\documentclass[12pt]{article}
\begin{document}

\vspace*{2cm} \noindent {\Large\bf Euler Walk on a Cayley Tree.}
\vspace{\baselineskip}
\newline {\bf A.E.\ Patrick\footnote[1]{Laboratory of Theoretical
Physics, Joint Institute for Nuclear Research, Dubna 141980,
Russia\\ e-mail: patrick@theor.jinr.ru}}
\begin{list}{}{\setlength{\rightmargin }{0mm}
\setlength{\leftmargin }{2.5cm}} \item \rule{124mm}{0.3mm}
\linebreak {\footnotesize {\bf Abstract.} We describe two possible
regimes (dynamic phases) of the Euler walk on a Cayley tree: a
condensed phase and a low-density phase. In the condensed phase
the area of visited sites grows as a compact domain. In the
low-density phase the proportion of visited sites decreases
rapidly from one generation of the tree to the next. We describe
in detail returns of the walker to the root and growth of the
domain of visited sites in the condensed phase. We also
investigate the critical behaviour of the model on the line
separating the two regimes.}
\newline \rule[1ex]{12.4cm}{0.3mm}
\linebreak {\bf \sc key words:} {\footnotesize Branching
processes; critical exponents; martingales; random walks.}
\vspace{\baselineskip}
\end{list}

\section{Introduction.} Consider a Cayley tree with arrows attached to
every site. Initially the arrows point at one of the
adjacent sites randomly and independently of each other, see Fig.\ 1. An
Eulerian walker moves over the Cayley tree according to the
following rules.  At time instants $l=0,1,2,\ldots$ the walker
jumps from its current location $x(l)$ (at one of the sites of the
tree) to the adjacent site in the direction of the arrow at
$x(l)$. At the time of jump the arrow at $x(l)$ is rotated
clockwise, till it points to another adjacent site.

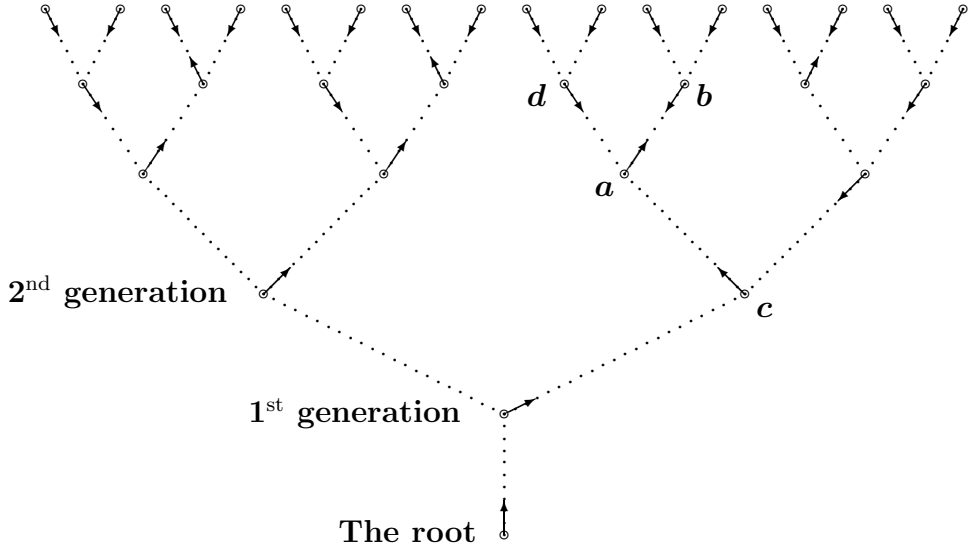
\begin{figure}[t]
\setlength{\unitlength}{1mm}
\begin{picture}(150,73)(0,-73)

\put(55,-71){\bf The root} \put(77,-70){\circle{1}}
\put(77,-70){\vector(0,1){4.6}}
\multiput(77.2,-70)(0,1.6){11}{\makebox(0,0){\circle*{0.2}}}

\put(43,-55){\bf 1${}^{\rm st}$ generation}

\put(77,-54){\vector(2,1){4.2}}
\multiput(77.2,-54)(1.46,0.73){22}{\makebox(0,0){\circle*{0.2}}}
\multiput(77.2,-54)(-1.46,0.73){22}{\makebox(0,0){\circle*{0.2}}}

\multiput(45.2,-38)(1,1){16}{\makebox(0,0){\circle*{0.2}}}
\multiput(45.2,-38)(-1,1){16}{\makebox(0,0){\circle*{0.2}}}

\multiput(109.2,-38)(1,1){16}{\makebox(0,0){\circle*{0.2}}}
\multiput(109.2,-38)(-1,1){16}{\makebox(0,0){\circle*{0.2}}}
\put(45,-38){\vector(1,1){3.6}} \put(109,-38){\vector(-1,1){3.6}}

\put(11,-39){\bf 2${}^{\rm nd}$ generation}

\multiput(29.2,-22)(1,1.5){8}{\makebox(0,0){\circle*{0.2}}}
\multiput(29.2,-22)(-1,1.5){8}{\makebox(0,0){\circle*{0.2}}}
\multiput(61.2,-22)(1,1.5){8}{\makebox(0,0){\circle*{0.2}}}
\multiput(61.2,-22)(-1,1.5){8}{\makebox(0,0){\circle*{0.2}}}
\multiput(93.2,-22)(1,1.5){8}{\makebox(0,0){\circle*{0.2}}}
\multiput(93.2,-22)(-1,1.5){8}{\makebox(0,0){\circle*{0.2}}}
\multiput(125.2,-22)(1,1.5){8}{\makebox(0,0){\circle*{0.2}}}
\multiput(125.2,-22)(-1,1.5){8}{\makebox(0,0){\circle*{0.2}}}
\put(29,-22){\vector(2,3){3}}\put(61,-22){\vector(2,3){3}}

\put(89,-25){\boldmath $a$}\put(110.5,-41){\boldmath
$c$}\put(102.5,-13){\boldmath $b$}\put(80,-13){\boldmath $d$}

\put(93,-22){\vector(2,3){3}}\put(125,-22){\vector(-1,-1){3.6}}

\multiput(21.2,-10)(0.72,1.44){7}{\makebox(0,0){\circle*{0.2}}}
\multiput(21.2,-10)(-0.72,1.44){7}{\makebox(0,0){\circle*{0.2}}}
\put(16,0){\vector(1,-2){1.8}} \put(26,0){\vector(-1,-2){1.8}}

\multiput(37.2,-10)(0.72,1.44){7}{\makebox(0,0){\circle*{0.2}}}
\multiput(37.2,-10)(-0.72,1.44){7}{\makebox(0,0){\circle*{0.2}}}
\put(32,0){\vector(1,-2){1.8}} \put(42,0){\vector(-1,-2){1.8}}

\multiput(53.2,-10)(0.72,1.44){7}{\makebox(0,0){\circle*{0.2}}}
\multiput(53.2,-10)(-0.72,1.44){7}{\makebox(0,0){\circle*{0.2}}}
\put(48,0){\vector(1,-2){1.8}} \put(58,0){\vector(-1,-2){1.8}}

\multiput(69.2,-10)(0.72,1.44){7}{\makebox(0,0){\circle*{0.2}}}
\multiput(69.2,-10)(-0.72,1.44){7}{\makebox(0,0){\circle*{0.2}}}
\put(64,0){\vector(1,-2){1.8}} \put(74,0){\vector(-1,-2){1.8}}

\multiput(85.2,-10)(0.72,1.44){7}{\makebox(0,0){\circle*{0.2}}}
\multiput(85.2,-10)(-0.72,1.44){7}{\makebox(0,0){\circle*{0.2}}}
\put(80,0){\vector(1,-2){1.8}} \put(90,0){\vector(-1,-2){1.8}}

\multiput(101.2,-10)(0.72,1.44){7}{\makebox(0,0){\circle*{0.2}}}
\multiput(101.2,-10)(-0.72,1.44){7}{\makebox(0,0){\circle*{0.2}}}
\put(96,0){\vector(1,-2){1.8}} \put(106,0){\vector(-1,-2){1.8}}

\multiput(117.2,-10)(0.72,1.44){7}{\makebox(0,0){\circle*{0.2}}}
\multiput(117.2,-10)(-0.72,1.44){7}{\makebox(0,0){\circle*{0.2}}}
\put(112,0){\vector(1,-2){1.8}} \put(122,0){\vector(-1,-2){1.8}}

\multiput(133.2,-10)(0.72,1.44){7}{\makebox(0,0){\circle*{0.2}}}
\multiput(133.2,-10)(-0.72,1.44){7}{\makebox(0,0){\circle*{0.2}}}
\put(128,0){\vector(1,-2){1.8}} \put(138,0){\vector(-1,-2){1.8}}

\put(77,-54){{\circle{1}}}
\multiput(45,-38)(64,0){2}{{\circle{1}}}
\multiput(29,-22)(32,0){4}{{\circle{1}}}
\multiput(21,-10)(16,0){8}{{\circle{1}}}
\multiput(16,0)(16,0){8}{{\circle{1}}}
\multiput(26,0)(16,0){8}{{\circle{1}}}

\multiput(16,0)(16,0){8}{{\circle*{0.2}}}
\multiput(26,0)(16,0){8}{{\circle*{0.2}}}

\put(21,-10){\vector(2,-3){2.5}} \put(37,-10){\vector(-1,2){1.8}}
\put(53,-10){\vector(2,-3){2.5}}\put(69,-10){\vector(-1,2){1.8}}
\put(85,-10){\vector(2,-3){2.5}}\put(101,-10){\vector(-2,-3){2.5}}
\put(117,-10){\vector(1,2){1.8}}\put(133,-10){\vector(-2,-3){2.5}}
\end{picture}
\caption{A Cayley tree with internal arrows arranged at random and
independently of each other. Boundary arrows always point inside
the tree to prevent the walker from falling off the tree. At site
$a$ the walker jumps in the direction of the arrow to site $b$,
and the arrow at $a$ is rotated clockwise to point at site $c$.
After that the walker jumps back to site $a$ and then jumps to
site $c$. At the time of the last jump the arrow at $a$ is again
rotated clockwise to point at site $d$.}
\end{figure}

We assume that the generations of the tree are numbered from
bottom to top. Zero generation of the tree contains only the root.
If $k>l$, then the generation number $k$ is above the generation
number $l$ on a picture of the tree, and we say that the $k^{\rm
th}$ generation is higher than the generation number $l$.

Eulerian walkers were introduced by Priezzhev et al \cite{p96}
(see \cite{p98} for further investigations) as a model of
``self-organized criticality". In their version of the model the
walker rotates arrows on arrival at a site of the graph. The two
versions are largely equivalent, but in the version used in this
paper it is easier to see on the picture of a graph where the
walker actually goes over the next few steps.

An attractive feature of the model on a {\em finite} graph with
reflecting boundary (the boundary arrows point inside the graph) is that eventually the walker settles into an
Euler circuit, where it passes every edge of the graph twice (once
in every direction). There is exactly one ``clockwise" Euler
circuit for any tree, see Fig.\ 2. As a result of the walker's
activity, initially chaotically oriented arrows arrange into an
organized configuration directing the walker around that circuit.

\begin{figure}[t]
\setlength{\unitlength}{1mm}
\begin{picture}(150,73)(0,-73)
\put(77,-70){\circle{1}} \put(77,-70){\vector(0,1){4}}
\multiput(77.2,-70)(0,1.6){11}{\makebox(0,0){\circle*{0.2}}}

\put(77,-54){\vector(-2,1){4}}
\multiput(77.2,-54)(1.46,0.73){22}{\makebox(0,0){\circle*{0.2}}}
\multiput(77.2,-54)(-1.46,0.73){22}{\makebox(0,0){\circle*{0.2}}}

\multiput(45.2,-38)(1,1){16}{\makebox(0,0){\circle*{0.2}}}
\multiput(45.2,-38)(-1,1){16}{\makebox(0,0){\circle*{0.2}}}

\multiput(109.2,-38)(1,1){16}{\makebox(0,0){\circle*{0.2}}}
\multiput(109.2,-38)(-1,1){16}{\makebox(0,0){\circle*{0.2}}}
\put(45,-38){\vector(-1,1){3.6}} \put(109,-38){\vector(-1,1){3.6}}

\multiput(29.2,-22)(1,1.5){8}{\makebox(0,0){\circle*{0.2}}}
\multiput(29.2,-22)(-1,1.5){8}{\makebox(0,0){\circle*{0.2}}}
\multiput(61.2,-22)(1,1.5){8}{\makebox(0,0){\circle*{0.2}}}
\multiput(61.2,-22)(-1,1.5){8}{\makebox(0,0){\circle*{0.2}}}
\multiput(93.2,-22)(1,1.5){8}{\makebox(0,0){\circle*{0.2}}}
\multiput(93.2,-22)(-1,1.5){8}{\makebox(0,0){\circle*{0.2}}}
\multiput(125.2,-22)(1,1.5){8}{\makebox(0,0){\circle*{0.2}}}
\multiput(125.2,-22)(-1,1.5){8}{\makebox(0,0){\circle*{0.2}}}

\put(29,-22){\vector(-2,3){3}}\put(61,-22){\vector(-2,3){3}}
\put(93,-22){\vector(-2,3){3}}\put(125,-22){\vector(-2,3){3}}

\multiput(21.2,-10)(0.72,1.44){7}{\makebox(0,0){\circle*{0.2}}}
\multiput(21.2,-10)(-0.72,1.44){7}{\makebox(0,0){\circle*{0.2}}}
\put(16,0){\vector(1,-2){1.8}} \put(26,0){\vector(-1,-2){1.8}}

\multiput(37.2,-10)(0.72,1.44){7}{\makebox(0,0){\circle*{0.2}}}
\multiput(37.2,-10)(-0.72,1.44){7}{\makebox(0,0){\circle*{0.2}}}
\put(32,0){\vector(1,-2){1.8}} \put(42,0){\vector(-1,-2){1.8}}

\multiput(53.2,-10)(0.72,1.44){7}{\makebox(0,0){\circle*{0.2}}}
\multiput(53.2,-10)(-0.72,1.44){7}{\makebox(0,0){\circle*{0.2}}}
\put(48,0){\vector(1,-2){1.8}} \put(58,0){\vector(-1,-2){1.8}}

\multiput(69.2,-10)(0.72,1.44){7}{\makebox(0,0){\circle*{0.2}}}
\multiput(69.2,-10)(-0.72,1.44){7}{\makebox(0,0){\circle*{0.2}}}
\put(64,0){\vector(1,-2){1.8}} \put(74,0){\vector(-1,-2){1.8}}

\multiput(85.2,-10)(0.72,1.44){7}{\makebox(0,0){\circle*{0.2}}}
\multiput(85.2,-10)(-0.72,1.44){7}{\makebox(0,0){\circle*{0.2}}}
\put(80,0){\vector(1,-2){1.8}} \put(90,0){\vector(-1,-2){1.8}}

\multiput(101.2,-10)(0.72,1.44){7}{\makebox(0,0){\circle*{0.2}}}
\multiput(101.2,-10)(-0.72,1.44){7}{\makebox(0,0){\circle*{0.2}}}
\put(96,0){\vector(1,-2){1.8}} \put(106,0){\vector(-1,-2){1.8}}

\multiput(117.2,-10)(0.72,1.44){7}{\makebox(0,0){\circle*{0.2}}}
\multiput(117.2,-10)(-0.72,1.44){7}{\makebox(0,0){\circle*{0.2}}}
\put(112,0){\vector(1,-2){1.8}} \put(122,0){\vector(-1,-2){1.8}}

\multiput(133.2,-10)(0.72,1.44){7}{\makebox(0,0){\circle*{0.2}}}
\multiput(133.2,-10)(-0.72,1.44){7}{\makebox(0,0){\circle*{0.2}}}
\put(128,0){\vector(1,-2){1.8}} \put(138,0){\vector(-1,-2){1.8}}

\put(77,-54){{\circle{1}}}

\multiput(45,-38)(64,0){2}{{\circle{1}}}
\multiput(29,-22)(32,0){4}{{\circle{1}}}
\multiput(21,-10)(16,0){8}{{\circle{1}}}
\multiput(16,0)(16,0){8}{{\circle{1}}}
\multiput(26,0)(16,0){8}{{\circle{1}}}

\multiput(16,0)(16,0){8}{{\circle*{0.2}}}
\multiput(26,0)(16,0){8}{{\circle*{0.2}}}

\put(21,-10){\vector(-1,2){1.8}} \put(37,-10){\vector(-1,2){1.8}}
\put(53,-10){\vector(-1,2){1.8}}\put(69,-10){\vector(-1,2){1.8}}
\put(85,-10){\vector(-1,2){1.8}}\put(101,-10){\vector(-1,2){1.8}}
\put(117,-10){\vector(-1,2){1.8}}\put(133,-10){\vector(-1,2){1.8}}

\thicklines %
\put(78,-70.25){\line(0,1){15.75}}
\put(76,-70.25){\line(0,1){15.75}}

\put(78,-54.5){\line(2,1){31.7}}\put(76,-54.5){\line(-2,1){31.7}}

\put(77,-53){\line(2,1){30.5}} \put(77,-53){\line(-2,1){30.5}}

\put(109.75,-38.6){\line(1,1){15.9}}
\put(107.4,-37.75){\line(-1,1){15}}
\put(46.6,-37.75){\line(1,1){15}}

\put(44.25,-38.6){\line(-1,1){15.9}}
\put(109.1,-36.6){\line(1,1){14.7}}
\put(109.1,-36.6){\line(-1,1){14.7}}
\put(45.1,-36.6){\line(1,1){14.7}}
\put(45.1,-36.6){\line(-1,1){14.7}}

\put(125.75,-22.6){\line(2,3){8}}
\put(94.25,-21.8){\line(2,3){7.75}}
\put(30.25,-21.8){\line(2,3){7.75}}
\put(59.85,-21.8){\line(-2,3){7.75}}
\put(123.85,-21.8){\line(-2,3){7.75}}

\put(28.25,-22.6){\line(-2,3){8}}
\put(92.25,-22.6){\line(-2,3){8}}
\put(61.75,-22.6){\line(2,3){8.15}}

\put(125,-20.6){\line(2,3){7}} \put(125,-20.6){\line(-2,3){7}}
\put(93,-20.6){\line(2,3){7}} \put(93,-20.6){\line(-2,3){7}}

\put(61,-20.6){\line(2,3){7}}
\put(61,-20.6){\line(-2,3){7}}\put(29,-20.6){\line(2,3){7}}
\put(29,-20.6){\line(-2,3){7}}

\put(133.75,-10.5){\line(1,2){5.7}}
\put(133,-8.0){\line(1,2){4.5}} \put(117,-8.0){\line(1,2){4.5}}
\put(101,-8.0){\line(1,2){4.5}}\put(85,-8.0){\line(1,2){4.5}}
\put(69,-8.0){\line(1,2){4.5}}\put(53,-8.0){\line(1,2){4.5}}
\put(37,-8.0){\line(1,2){4.5}}\put(21,-8.0){\line(1,2){4.5}}

\put(132,-10){\line(-1,2){5.5}}

\put(118,-10){\line(1,2){5.5}} \put(116,-10){\line(-1,2){5.5}}
\put(100,-10){\line(-1,2){5.5}} \put(102,-10){\line(1,2){5.5}}
\put(84.25,-10.55){\line(-1,2){5.75}}

\put(86,-10){\line(1,2){5.5}} \put(68,-10){\line(-1,2){5.5}}
\put(69.85,-10.45){\line(1,2){5.7}} \put(52,-10){\line(-1,2){5.5}}
\put(54,-10){\line(1,2){5.5}} \put(36,-10){\line(-1,2){5.5}}
\put(38,-10){\line(1,2){5.5}}
\put(20.25,-10.55){\line(-1,2){5.75}}
\put(22,-10){\line(1,2){5.5}}

\put(133,-8.0){\line(-1,2){4.5}} \put(117,-8.0){\line(-1,2){4.5}}
\put(101,-8.0){\line(-1,2){4.5}}\put(85,-8.0){\line(-1,2){4.5}}
\put(69,-8.0){\line(-1,2){4.5}} \put(53,-8.0){\line(-1,2){4.5}}
\put(37,-8.0){\line(-1,2){4.5}} \put(21,-8.0){\line(-1,2){4.5}}

\multiput(14.4,1)(16,0){8}{\line(1,0){2.2}}
\multiput(25.4,1)(16,0){8}{\line(1,0){2.2}}
\end{picture}
\caption{The Euler circuit on a Cayley tree. The orientation of
arrows corresponds to the current position of the walker at the
root of the tree.}
\end{figure}
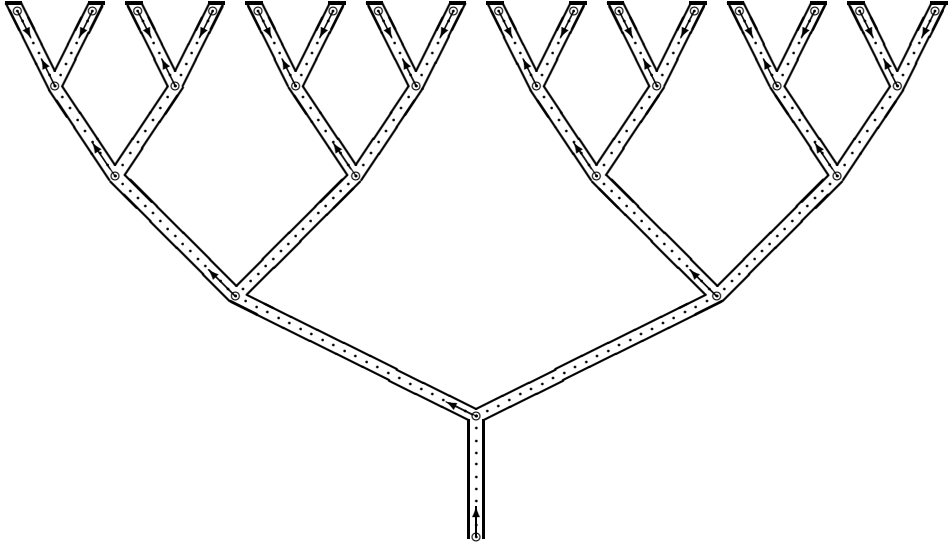

If the Euler circuit on a tree is a critical state is not that
clear. It is not difficult to calculate correlation functions for
orientations of the arrows at two sites of a Cayley tree, assuming
the uniform distribution of the current location of the walker.
For instance, let $a_1$ and $a_2$ be two arrows at sites in the
generations $k$ and $k+m$ of a finite tree containing $n$
generations in total. Then
\[
\Pr[a_1=\searrow,a_2=\searrow]-\Pr[a_1=\searrow]\Pr[a_2=\searrow]\to
-2^{-2k-m},\quad\mbox{as }n\to\infty.
\]
Hence, we have an exponential decay of correlations with the
distance between the arrows measured in generations of the tree.
The same asymptotic behaviour we obtain for all other correlation
functions, although some of those are positive.

The above decay of correlations is in contrast to the behaviour
found by Dhar and Majumdar for the self-organized state of a sand
pile on a Cayley tree, see \cite{dm90}. Dhar and Majumdar found
that the correlation functions decay as $4^{-m}$, where $m$ is the
distance between the two sites of the tree. They concluded that
the correlations are short-ranged, because even after
multiplication by the branching factor $2^m$ one still has an
exponential decay to 0. Nevertheless they classified the
self-organized state of the sand pile on a Cayley tree as
critical, presumably, because of power-law tails in the
distributions of avalanche-sizes and related quantities.

In our case the correlations decay as $2^{-m}$, and do not vanish
after multiplication by the branching factor. Therefore one
certainly can not rule out the criticality of the Euler circuit on
a tree on the basis of exponential decay of correlation functions.
Nevertheless one feels that the self-organized state in this case
is closer to the minimally stable state of the 1D sand pile,
described in the paper \cite{btw88}, than to a truly critical
state.

Our main goal in this paper is a description of the {\em
formation} of an organized structure on an infinite tree. We will
show that, unlike what one sees on finite graphs with reflecting
boundary, on an infinite tree a (substantial density of) organized
structure is not always formed. Of course, if an organized
structure is not formed on an infinite tree, it is highly
sensitive to the boundary conditions and appears on a finite tree
only as a result of numerous bounces of the walker against the
reflecting boundary.

To set the scenery for the study of Euler walk let us describe two
possible regimes of evolution: a condensed phase and a low-density
phase. By (dynamic) phases in this paper we mean not a particular
distribution $P[\mbox{\boldmath $a$}]$ of arrows {\boldmath $a$},
but a particular {\em type of evolution} of those distributions
$P_t[\mbox{\boldmath $a$}]$.

To describe the condensed phase let us arrange all the arrows
(except the one at the root of the tree) downwards, along the
edges of the tree. In this case the walker starting at the root at
time $T_0=0$ returns to the root at time instants $T_1=2$,
$T_2=8$, $T_3=22,\ldots$. In general, the $f^{\rm th}$ return to
the root takes place at the time instant $T_f=2^{f+2}-2f-4$.

There is a growing domain of visited sites --- the explored domain
--- which penetrates the $k^{\rm th}$ generation of the tree at the
time instant $t_k\equiv 2^{k+1}-k-2$, $k=1,2,\ldots$. At the time
instant $s_k\equiv 2^{k+2}-3k-3$ the domain swallows the $k^{\rm
th}$ generation completely, and the walker heads toward the root.

If we denote $g_{\rm max}(t)$ the highest generation visited by
the walker by the time $t$, then the formula for $t_k$ yields
\[
\log_2(t)-1\leq g_{\rm max}(t)\leq\log_2(t),\quad\mbox{for
}t\geq4.
\]
Analogously, if we denote $g_{\rm c}(t)$ the number of generations
completely explored by the time $t$, then the formula for $s_k$
yields
\[
\log_2(t)-2\leq g_{\rm c}(t)\leq\log_2(t)-1,\quad\mbox{for
}t\geq3.
\]
Thus, for the downward initial arrangement of the arrows, the
growing explored area is a ``compact" domain of the tree. The
height of the domain (measured in generations) grows with time as
$\log_2t$. Below the highest visited generation the density of
visited sites is 1, above that generation the density of visited
sites is, of course, 0. This is the {\em condensed phase} of the
Euler walk.

Another regime --- {\em the low-density phase} --- is obtained if
we begin with the upward (left or right) initial orientation of
the arrows. In this case the walker goes straight toward the top
of the tree. The density of visited sites in the $f^{\rm th}$
generation at time $t$ is $2^{-f+1}$ (for $f\leq t$), which tends
to $0$ with $f$ justifying the name the low-density phase. Of
course, once the walker reaches the top of the (finite) tree it
turns back and gradually stomps the whole graph. Therefore (as it
should be) a clear-cut distinction between the two phases exists
only on an infinite tree.

For a random initial arrangements of the arrows we obtain a phase
which is a perturbation of either the condensed or the low-density
phase. As we will see in the following sections, the transition
between the two phases takes place when
$2\Pr[\nwarrow]+\Pr[\nearrow]=1$.

The rest of the paper is organized as follows. In Section 2 we
investigate the properties of the condensed phase: the returns of
the walker to the root, and the growth of the explored domain when
$2\Pr[\nwarrow]+\Pr[\nearrow]<1$. In Section 3 we show the absence
of a compact domain of visited sites if
$2\Pr[\nwarrow]+\Pr[\nearrow]>1$. In Section 4 we repeat the
program of Section 2 at the critical point
$2\Pr[\nwarrow]+\Pr[\nearrow]=1$. Traditionally, the last section
is devoted to a discussion of the results obtained in the previous
sections.

\section{The condensed phase.}
Let the internal arrows be initially arranged independently of one
another, and according to the distribution $\Pr[\nwarrow]=p$,
$\Pr[\nearrow]=q$, $\Pr[\downarrow]=1-p-q$. One can map every
initial configuration of arrows into a realization of a
discrete-time branching process according to the following rules.

Place a particle at the root of the tree. This particle produces
exactly one descendant --- a particle which is placed at the site
of the first generation of the tree. From the first generation on,
a particle produces either 0, or 1, or 2 descendants depending on
the initial direction of the arrow at the site occupied by the
particle. If the arrow points downward, then the particle does not
have descendants. If the arrow points up and right (like the arrow
at the first-generation site on Fig.\ 3), then the particle has
exactly one descendant placed at the adjacent site in the
direction of the arrow. Finally, if the arrow points up and left,
then there are exactly two descendants placed at the two adjacent
sites above, see Fig.\ 3.

\begin{figure}[t]
\setlength{\unitlength}{1mm}
\begin{picture}(150,73)(0,-73)
\put(77,-70){\circle*{1}} \put(77,-70){\vector(0,1){4.6}}
\multiput(77.2,-70)(0,1.6){11}{\makebox(0,0){\circle*{0.2}}}
\put(77,-54){\vector(2,1){4.2}}
\multiput(77.2,-54)(1.46,0.73){22}{\makebox(0,0){\circle*{0.2}}}
\multiput(77.2,-54)(-1.46,0.73){22}{\makebox(0,0){\circle*{0.2}}}
\multiput(45.2,-38)(1,1){16}{\makebox(0,0){\circle*{0.2}}}
\multiput(45.2,-38)(-1,1){16}{\makebox(0,0){\circle*{0.2}}}
\multiput(109.2,-38)(1,1){16}{\makebox(0,0){\circle*{0.2}}}
\multiput(109.2,-38)(-1,1){16}{\makebox(0,0){\circle*{0.2}}}
\put(45,-38){\vector(1,1){3.6}} \put(109,-38){\vector(-1,1){3.6}}

\multiput(29.2,-22)(1,1.5){8}{\makebox(0,0){\circle*{0.2}}}
\multiput(29.2,-22)(-1,1.5){8}{\makebox(0,0){\circle*{0.2}}}
\multiput(61.2,-22)(1,1.5){8}{\makebox(0,0){\circle*{0.2}}}
\multiput(61.2,-22)(-1,1.5){8}{\makebox(0,0){\circle*{0.2}}}
\multiput(93.2,-22)(1,1.5){8}{\makebox(0,0){\circle*{0.2}}}
\multiput(93.2,-22)(-1,1.5){8}{\makebox(0,0){\circle*{0.2}}}
\multiput(125.2,-22)(1,1.5){8}{\makebox(0,0){\circle*{0.2}}}
\multiput(125.2,-22)(-1,1.5){8}{\makebox(0,0){\circle*{0.2}}}
\put(29,-22){\vector(2,3){3}}\put(61,-22){\vector(2,3){3}}
\put(93,-22){\vector(2,3){3}}\put(125,-22){\vector(-1,-1){3.6}}

\multiput(21.2,-10)(0.72,1.44){7}{\makebox(0,0){\circle*{0.2}}}
\multiput(21.2,-10)(-0.72,1.44){7}{\makebox(0,0){\circle*{0.2}}}
\put(16,0){\vector(1,-2){1.8}} \put(26,0){\vector(-1,-2){1.8}}
\multiput(37.2,-10)(0.72,1.44){7}{\makebox(0,0){\circle*{0.2}}}
\multiput(37.2,-10)(-0.72,1.44){7}{\makebox(0,0){\circle*{0.2}}}
\put(32,0){\vector(1,-2){1.8}} \put(42,0){\vector(-1,-2){1.8}}
\multiput(53.2,-10)(0.72,1.44){7}{\makebox(0,0){\circle*{0.2}}}
\multiput(53.2,-10)(-0.72,1.44){7}{\makebox(0,0){\circle*{0.2}}}
\put(48,0){\vector(1,-2){1.8}} \put(58,0){\vector(-1,-2){1.8}}
\multiput(69.2,-10)(0.72,1.44){7}{\makebox(0,0){\circle*{0.2}}}
\multiput(69.2,-10)(-0.72,1.44){7}{\makebox(0,0){\circle*{0.2}}}
\put(64,0){\vector(1,-2){1.8}} \put(74,0){\vector(-1,-2){1.8}}
\multiput(85.2,-10)(0.72,1.44){7}{\makebox(0,0){\circle*{0.2}}}
\multiput(85.2,-10)(-0.72,1.44){7}{\makebox(0,0){\circle*{0.2}}}
\put(80,0){\vector(1,-2){1.8}} \put(90,0){\vector(-1,-2){1.8}}
\multiput(101.2,-10)(0.72,1.44){7}{\makebox(0,0){\circle*{0.2}}}
\multiput(101.2,-10)(-0.72,1.44){7}{\makebox(0,0){\circle*{0.2}}}
\put(96,0){\vector(1,-2){1.8}} \put(106,0){\vector(-1,-2){1.8}}
\multiput(117.2,-10)(0.72,1.44){7}{\makebox(0,0){\circle*{0.2}}}
\multiput(117.2,-10)(-0.72,1.44){7}{\makebox(0,0){\circle*{0.2}}}
\put(112,0){\vector(1,-2){1.8}} \put(122,0){\vector(-1,-2){1.8}}
\multiput(133.2,-10)(0.72,1.44){7}{\makebox(0,0){\circle*{0.2}}}
\multiput(133.2,-10)(-0.72,1.44){7}{\makebox(0,0){\circle*{0.2}}}
\put(128,0){\vector(1,-2){1.8}} \put(138,0){\vector(-1,-2){1.8}}

\put(77,-54){\circle*{1}} \put(73,-52.5){${}_\spadesuit$}
\put(109,-38){\circle*{1}} \multiput(45,-38)(64,0){2}{\circle{1}}
\multiput(93,-22)(32,0){2}{\circle*{1}}
\multiput(29,-22)(32,0){4}{{\circle{1}}}
\put(122,-19.3){${}_\spadesuit$} \put(126,-19.3){${}_\spadesuit$}
\put(90,-19.3){${}_\spadesuit$}
\multiput(21,-10)(16,0){8}{{\circle{1}}}
\put(101,-10){\circle*{1}} \put(98.7,-7.2){${}_\spadesuit$}
\put(101.3,-7.2){${}_\spadesuit$}
\multiput(16,0)(16,0){8}{{\circle{1}}}
\multiput(26,0)(16,0){8}{{\circle{1}}}

\multiput(16,0)(16,0){8}{{\circle*{0.2}}}
\multiput(26,0)(16,0){8}{{\circle*{0.2}}}

\put(21,-10){\vector(2,-3){2.5}} \put(37,-10){\vector(-1,2){1.8}}
\put(53,-10){\vector(2,-3){2.5}}\put(69,-10){\vector(-2,-3){2.5}}
\put(85,-10){\vector(1,2){1.8}}\put(101,-10){\vector(-2,-3){2.5}}
\put(117,-10){\vector(1,2){1.8}}\put(133,-10){\vector(-2,-3){2.5}}

\thicklines %
\put(76,-70.25){\line(0,1){16.8}}
\put(78,-70.25){\line(0,1){15.75}}

\put(78,-54.55){\line(2,1){31.7}} \put(76,-53.5){\line(2,1){31.5}}

\put(107.5,-37.75){\line(-1,1){15.6}}
\put(109,-36.6){\line(-1,1){14.7}}

\put(109.75,-38.6){\line(1,1){17.5}}
\put(109.1,-36.6){\line(1,1){15.6}}

\put(124.6,-21){\line(1,0){2.8}}

\put(94.3,-21.89){\line(2,3){8.5}}
\put(91.9,-22.05){\line(2,3){8.6}} \put(100.5,-9){\line(1,0){2.4}}
\end{picture}
\caption{An initial arrangement of the arrows, the corresponding
first return to the root of the Euler walker (solid lines), the
first-return cluster of the associated branching process (discs),
and the buds (spades). At the next visit to a site with buds an
independent first-return cluster will grow from every bud.}
\end{figure}

The relevance of the branching process to our main problem stems
from the following fact. If the branching process degenerates,
then the walker returns to the root at a finite time-instant $T_1$
equal twice the number of descendants in the branching process
(not counting the original particle at the root). The first-return
path encircles the particles in all generations of the branching
process, which we call below {\em the first-return cluster}.

The above correspondence between paths of the walker on Cayley
tree and realizations of the branching process allows one to
employ the elegant technique of generating functions and the main
results from the theory of branching processes \cite{h91,f70}.
First of all recall that if a particle produces $k$ descendants
with probability $p_k$, then the branching process degenerates
with probability 1 if and only if $\sum_{k=1}^\infty kp_k\leq 1$.
Hence, the time of the first return is finite with probability 1
if and only if $q+2p\leq 1$. The critical case $q+2p=1$ requires a
special consideration, therefore, in this section we consider only
the case $q+2p<1$.

{\sf Lemma 1.} Let $q+2p<1$, then the walker returns to the root
for the first time at an almost surely finite even time-instant
$T_1$, such that
\begin{eqnarray}
&&m_1\equiv\mbox{\boldmath $E$}T_1=\frac{2}{1-(q+2p)};\nonumber\\
&&\mbox{Var}\,T_1=\frac{4(1-q)}{(1-(q+2p))^3}-\frac{4}{1-(q+2p)};\nonumber
\end{eqnarray}
\[
\Pr[T_1=2k]\sim \sqrt{\frac{q\sqrt{(1-p-q)/p}+2(1-p-q)}{4\pi p}}\,
k^{-3/2}\left(q+2\sqrt{p(1-p-q)}\right)^k,
\]
as $k\to\infty$.

{\sf Proof.} Denote $X$ the number of descendants for a particle
outside the root of the tree. The probability generating function
of $X$ is given by
\begin{equation}
g(y)\equiv\mbox{\boldmath $E$}y^X=1-q-p+qy+py^2.
\label{gfg}
\end{equation}
Denote $Z$ the total number of descendants in the associated
branching process. The probability generating function of $Z$,
$f(x)\equiv\mbox{\boldmath $E$}x^Z$, is a solution of the
equation, see \cite{h91,f70},
\[
f(x)=xg(f(x)).
\]
Hence
\[
f(x)=\frac{1}{2px}\left[1-qx-\sqrt{(1-qx)^2-4p(1-p-q)x^2}\right].
\]
Differentiating $f(x)$ and taking into account $T_1=2Z$, we obtain
\[
m_1\equiv\mbox{\boldmath $E$}T_1=\frac{2}{1-(q+2p)},\quad
\mbox{Var}\,T_1=\frac{4(1-q)}{(1-(q+2p))^3}-\frac{4}{1-(q+2p)}.
\]
The above generating function $f(x)=\sum_{k=0}^\infty p_k x^k$
often appears in the literature on branching processes, see, e.g.,
the paper \cite{o49} by Otter. In particular, it is shown in that
paper that the large-$k$ asymptotics for $p_k=P[Z=k]$ is given by
\begin{equation}
p_k\sim \sqrt{\frac{q\sqrt{(1-p-q)/p}+2(1-p-q)}{4\pi
p}}\,k^{-3/2}\left(q+2\sqrt{p(1-p-q)}\right)^k, \label{pas}
\end{equation}
which is the announced formula for $P[T_1=2k]$ in the statement of
this lemma.

\hfill \rule{2mm}{2mm}\newline

{\sf Remark 1.} Denote $Z_k$ the number of particles of the
associated branching process in the $k^{\rm th}$ generation of the
tree. The random variable $X$ is the number of descendants
produced by a single particle as in the proof of Lemma 1. Then the
distribution of the height of the first-return path, $H_1$, is
given by
\[
\Pr[H_1=k]=\Pr[Z_{k+1}=0]-\Pr[Z_k= 0].
\]
It is shown in the book by Harris \cite{h91} that the large-$k$
asymptotics of $\Pr[Z_k= 0]$ is given by
\[
\Pr[Z_k=0]\sim 1-c_1(\mbox{\boldmath $E$}X)^k,
\]
if $\mbox{\boldmath $E$}X<1$, where $c_1$ is an unknown positive
constant.

Hence in our case the distribution of $H_1$ decays exponentially
with $k$,
\[
\Pr[H_1=k]\sim c(q+2p)^k.
\]
\hfill \rule{2mm}{2mm}

Thus, during the first stage of exploration of the Cayley tree
($0\leq t\leq T_1$) the walker stomps a first-return path with
statistical properties described in Lemma 1. To visualize the
motion of the walker after the first return to the root one can
imagine that, whenever a site is visited for the first time and
$X$ descendants are produced in the associated branching process,
the walker attaches $2-X$ buds to the site, see Fig.\ 3. During
the second stage of exploration (after the first return but before
the second return to the root) the walker follows the first-return
path, but, whenever a bud is encountered, it wonders off the
beaten track and appends to the existing path a new circuit, which
(unless hitting the boundary) is statistically equivalent to the
first-return path, see Fig.\ 4.

{\sf Lemma 2.} Let the walker return to the root for the first
time at time $T_1$. Then the first-return path has exactly
$1+\frac{1}{2}T_1$ attached buds.

{\sf Proof.} Recall the following standard representation for the
number of descendants, $Z_k$, in generations $k=2,3,\ldots$ of the
associated branching process
\begin{eqnarray}
&&Z_2=X_1^{(1)},\nonumber\\
&&Z_3=X_1^{(2)}+X_2^{(2)}+\ldots+X_{Z_2}^{(2)},\nonumber\\
&&Z_4=X_1^{(3)}+X_2^{(3)}+\ldots+X_{Z_3}^{(3)},\nonumber\\
&&\mbox{and so on,}\nonumber
\end{eqnarray}
where $X_k^{(l)}$ is the number of descendants produced by the
$k^{\rm th}$ particle from the $l^{\rm th}$ generation. All the
random variables $X_k^{(l)}$ are independent and have the same
distribution as the random variable $X$. Note also that $Z_1=1$,
and $Z_{n+1}=0$ whenever $Z_n=0$.

Then we have the following formulae for the number of buds $b_k$,
in generations $k=1,2,3,\ldots$
\begin{eqnarray}
&&b_1=2-X_1^{(1)}=2-Z_2,\nonumber\\
&&b_2=2-X_1^{(2)}+2-X_2^{(2)}+\ldots+2-X_{Z_2}^{(2)}=2Z_2-Z_3,\nonumber\\
&&b_3=2-X_1^{(3)}+2-X_2^{(3)}+\ldots+2-X_{Z_3}^{(3)}=2Z_3-Z_4,\nonumber\\
&&\mbox{and so on.}\nonumber
\end{eqnarray}
Since for $q+2p<1$ only a finite number of $Z_k$ have non-zero
values, the total number of buds on the first-return path is given
by
\[
B_1=\sum_{k=1}^\infty b_k=\sum_{k=2}^\infty
(2Z_{k-1}-Z_k)=1+\sum_{k=1}^\infty Z_k.
\]
The total number of descendants in all generations is
$\frac{1}{2}T_1$, hence $B_1=1+\frac{1}{2}T_1$.

\hfill \rule{2mm}{2mm}\newline
\begin{figure}[t]
\setlength{\unitlength}{1mm}
\begin{picture}(150,73)(0,-73)
\put(77,-70){\circle*{1}} \put(77,-70){\vector(0,1){4.6}}
\multiput(77.2,-70)(0,1.6){11}{\makebox(0,0){\circle*{0.2}}}
\put(77,-54){\vector(2,1){4.2}}
\multiput(77.2,-54)(1.46,0.73){22}{\makebox(0,0){\circle*{0.2}}}
\multiput(77.2,-54)(-1.46,0.73){22}{\makebox(0,0){\circle*{0.2}}}
\multiput(45.2,-38)(1,1){16}{\makebox(0,0){\circle*{0.2}}}
\multiput(45.2,-38)(-1,1){16}{\makebox(0,0){\circle*{0.2}}}
\multiput(109.2,-38)(1,1){16}{\makebox(0,0){\circle*{0.2}}}
\multiput(109.2,-38)(-1,1){16}{\makebox(0,0){\circle*{0.2}}}
\put(45,-38){\vector(1,1){3.6}} \put(109,-38){\vector(-1,1){3.6}}

\multiput(29.2,-22)(1,1.5){8}{\makebox(0,0){\circle*{0.2}}}
\multiput(29.2,-22)(-1,1.5){8}{\makebox(0,0){\circle*{0.2}}}
\multiput(61.2,-22)(1,1.5){8}{\makebox(0,0){\circle*{0.2}}}
\multiput(61.2,-22)(-1,1.5){8}{\makebox(0,0){\circle*{0.2}}}
\multiput(93.2,-22)(1,1.5){8}{\makebox(0,0){\circle*{0.2}}}
\multiput(93.2,-22)(-1,1.5){8}{\makebox(0,0){\circle*{0.2}}}
\multiput(125.2,-22)(1,1.5){8}{\makebox(0,0){\circle*{0.2}}}
\multiput(125.2,-22)(-1,1.5){8}{\makebox(0,0){\circle*{0.2}}}
\put(29,-22){\vector(2,3){3}}\put(61,-22){\vector(2,3){3}}
\put(93,-22){\vector(2,3){3}}\put(125,-22){\vector(-1,-1){3.6}}

\multiput(21.2,-10)(0.72,1.44){7}{\makebox(0,0){\circle*{0.2}}}
\multiput(21.2,-10)(-0.72,1.44){7}{\makebox(0,0){\circle*{0.2}}}
\put(16,0){\vector(1,-2){1.8}} \put(26,0){\vector(-1,-2){1.8}}
\multiput(37.2,-10)(0.72,1.44){7}{\makebox(0,0){\circle*{0.2}}}
\multiput(37.2,-10)(-0.72,1.44){7}{\makebox(0,0){\circle*{0.2}}}
\put(32,0){\vector(1,-2){1.8}} \put(42,0){\vector(-1,-2){1.8}}
\multiput(53.2,-10)(0.72,1.44){7}{\makebox(0,0){\circle*{0.2}}}
\multiput(53.2,-10)(-0.72,1.44){7}{\makebox(0,0){\circle*{0.2}}}
\put(48,0){\vector(1,-2){1.8}} \put(58,0){\vector(-1,-2){1.8}}
\multiput(69.2,-10)(0.72,1.44){7}{\makebox(0,0){\circle*{0.2}}}
\multiput(69.2,-10)(-0.72,1.44){7}{\makebox(0,0){\circle*{0.2}}}
\put(64,0){\vector(1,-2){1.8}} \put(74,0){\vector(-1,-2){1.8}}
\multiput(85.2,-10)(0.72,1.44){7}{\makebox(0,0){\circle*{0.2}}}
\multiput(85.2,-10)(-0.72,1.44){7}{\makebox(0,0){\circle*{0.2}}}
\put(80,0){\vector(1,-2){1.8}} \put(90,0){\vector(-1,-2){1.8}}
\multiput(101.2,-10)(0.72,1.44){7}{\makebox(0,0){\circle*{0.2}}}
\multiput(101.2,-10)(-0.72,1.44){7}{\makebox(0,0){\circle*{0.2}}}
\put(96,0){\vector(1,-2){1.8}} \put(106,0){\vector(-1,-2){1.8}}
\multiput(117.2,-10)(0.72,1.44){7}{\makebox(0,0){\circle*{0.2}}}
\multiput(117.2,-10)(-0.72,1.44){7}{\makebox(0,0){\circle*{0.2}}}
\put(112,0){\vector(1,-2){1.8}} \put(122,0){\vector(-1,-2){1.8}}
\multiput(133.2,-10)(0.72,1.44){7}{\makebox(0,0){\circle*{0.2}}}
\multiput(133.2,-10)(-0.72,1.44){7}{\makebox(0,0){\circle*{0.2}}}
\put(128,0){\vector(1,-2){1.8}} \put(138,0){\vector(-1,-2){1.8}}

\put(77,-54){\circle*{1}} \put(42.2,-36){${}_\spadesuit$}
\put(109,-38){\circle*{1}}

\multiput(45,-38)(64,0){2}{\circle{1}}
\multiput(93,-22)(32,0){2}{\circle*{1}}
\multiput(29,-22)(32,0){4}{{\circle{1}}}
\put(58.2,-19.5){${}_\spadesuit$}

\put(133.3,-7.5){${}_\spadesuit$}\put(130.6,-7.5){${}_\spadesuit$}

\put(85.2,-7.5){${}_\spadesuit$}\put(82.6,-7.5){${}_\spadesuit$}

\multiput(21,-10)(16,0){8}{{\circle{1}}}
\put(101,-10){\circle*{1}}
\put(114.5,-7.5){${}_\spadesuit$}
\multiput(16,0)(16,0){8}{{\circle{1}}}
\multiput(26,0)(16,0){8}{{\circle{1}}}

\multiput(16,0)(16,0){8}{{\circle*{0.2}}}
\multiput(26,0)(16,0){8}{{\circle*{0.2}}}

\put(21,-10){\vector(2,-3){2.5}} \put(37,-10){\vector(-1,2){1.8}}
\put(53,-10){\vector(2,-3){2.5}}\put(69,-10){\vector(-1,2){1.8}}
\put(85,-10){\vector(2,-3){2.5}}\put(101,-10){\vector(-2,-3){2.5}}
\put(117,-10){\vector(1,2){1.8}}\put(133,-10){\vector(-2,-3){2.5}}

\thicklines %
\put(78,-70.25){\line(0,1){15.75}}
\put(76,-70.25){\line(0,1){15.75}}

\put(78,-54.6){\line(2,1){31.7}} \put(76,-54.6){\line(-2,1){32.6}}
\put(77,-52.9){\line(-2,1){30.4}} \put(77,-52.9){\line(2,1){30.3}}

\put(109.8,-38.6){\line(1,1){15.8}}
\put(109,-36.6){\line(1,1){14.8}}

\put(46.6,-37.75){\line(1,1){15.3}}
\put(43.4,-38.2){\line(1,1){16.6}}
\put(109,-36.6){\line(-1,1){14.7}}
\put(107.4,-37.75){\line(-1,1){15.3}}

\put(125.75,-22.7){\line(2,3){9.1}}
\put(125,-20.3){\line(2,3){7.5}}

\put(125,-20.3){\line(-2,3){6.9}}
\put(123.8,-21.8){\line(-2,3){7.8}}

\put(94.3,-21.85){\line(2,3){8}} \put(93,-20.2){\line(2,3){6.9}}

\put(93,-20.2){\line(-2,3){7.4}}
\put(92.1,-22.3){\line(-2,3){8.8}}

\put(62,-22.4){\line(2,3){8.1}} \put(60,-21.6){\line(2,3){7.9}}
\put(118.1,-10){\line(1,2){5.5}} \put(116,-10){\line(1,2){5.5}}

\put(102,-10.2){\line(1,2){5.6}} \put(101,-7.8){\line(1,2){4.4}}
\put(101,-7.8){\line(-1,2){4.4}} \put(99.8,-9.8){\line(-1,2){5.4}}

\put(70,-10.2){\line(1,2){5.6}} \put(69,-7.8){\line(1,2){4.4}}
\put(69,-7.8){\line(-1,2){4.4}} \put(67.8,-9.7){\line(-1,2){5.3}}

\put(132.35,-9){\line(1,0){2.6}} \put(83.2,-9){\line(1,0){2.5}}

\put(121.4,1){\line(1,0){2.25}} \put(105.3,1){\line(1,0){2.3}}
\put(94.5,1){\line(1,0){2.2}} \put(73.2,1){\line(1,0){2.3}}
\put(62.3,1){\line(1,0){2.3}}

\end{picture}
\caption{The initial arrangement of the arrows, the corresponding
second return to the root for the Euler walker (solid lines), and
a new set of buds (spades).}
\end{figure}
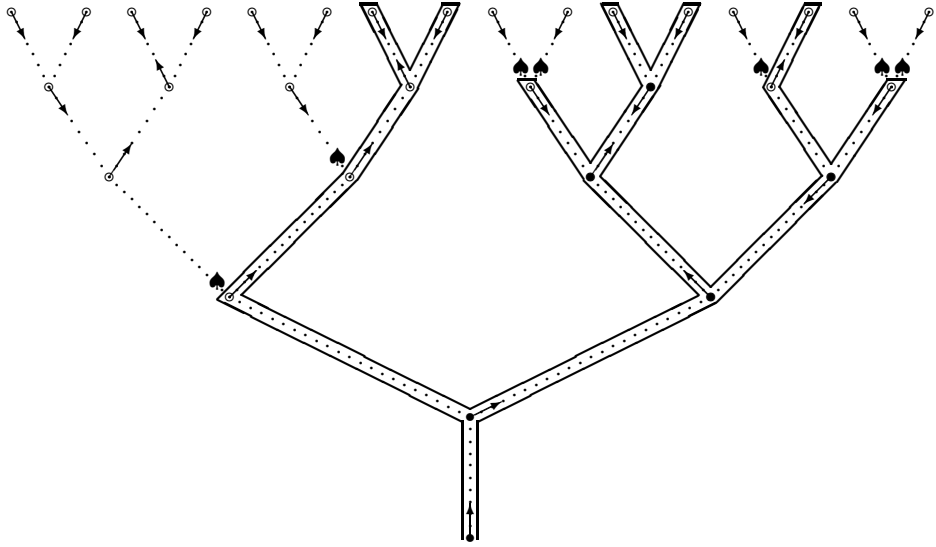

{\sf Theorem 1.} Let $q+2p<1$, then the Euler walker returns to
the root infinitely often at (almost surely finite) time instants
$T_1,T_2,T_3,\ldots$. Moreover, the sequence of normalized
differences
\[
Y_n=\frac{T_n-T_{n-1}+2}{(1+\frac{1}{2}\mbox{\boldmath
$E$}T_1)^n},\quad n=1,2,3,\ldots,
\]
is a positive and uniformly integrable martingale,
$\mbox{\boldmath $E$}[Y_n|Y_{n-1},\ldots,Y_1]=Y_{n-1}$.

{\sf Proof.} In order to return to the root for the second time
the walker has to repeat the first-return path and to create new
first-return circuits at each of the $B_1$ buds. Hence for the
time of the second return to the root we obtain
\[
T_2-T_1=T_1-T_0+\tau_1^{(2)}+\tau_2^{(2)}+\ldots+\tau_{B_1}^{(2)},
\]
where $T_0=0$, and $\tau_j^{(2)}$ are independent random variables
with the same distribution as the first-return time $T_1$. A
verbatim repetition of the argument from the proof of Lemma 2
shows that on each of the new circuits attached to the
first-return path the walker creates $1+\frac{1}{2}\tau_j^{(2)}$
buds, $j=1,2,\ldots,B_1$. Hence the total number of buds on the
second-return path is given by
\[
B_2=B_1+\frac{1}{2}\sum_{j=1}^{B_1}\tau_j^{(2)}.
\]

We have essentially the same scenario for any return to the root.
For the time of the $n^{\rm th}$ return to the root we obtain
\begin{equation}
T_n-T_{n-1}=T_{n-1}-T_{n-2}+\sum_{j=1}^{B_{n-1}}\tau_j^{(n)}.
\label{tn}
\end{equation}
The number of buds on the $n^{\rm th}$ return path is given by
\[
B_n=B_{n-1}+\frac{1}{2}\sum_{j=1}^{B_{n-1}}\tau_j^{(n)}.
\]
Since $B_1=1+\frac{1}{2}T_1$, we can rewrite the last equation as
\begin{equation}
B_n=\sum_{j=1}^{B_{n-1}}b_j^{(n)}. \label{bn}
\end{equation}
where $b_j^{(n)}=1+\frac{1}{2}\tau_j^{(n)}$ are independent random
variables with the same distribution as $B_1$.

Induction and the obtained relationships for $T_n$ and $B_n$ yield
\begin{equation}
B_n=1+{\textstyle\frac{1}{2}}(T_n-T_{n-1}),\quad\mbox{for any
}n\geq1. \label{bntn}
\end{equation}
Indeed, Lemma 2 says that in the case $n=1$ this formula is
correct. Suppose that the formula is also correct for $n=k$. Then
the relationships for $B_n$ and $T_n$ yield
\begin{eqnarray}
B_{k+1}=B_{k}+\frac{1}{2}\sum_{j=1}^{B_{k}}\tau_j^{(k)}&=&
1+{\textstyle\frac{1}{2}}(T_k-T_{k-1})+{\textstyle\frac{1}{2}}
(T_{k+1}-T_{k}-T_{k}+T_{k-1})\nonumber\\&=&
1+{\textstyle\frac{1}{2}}(T_{k+1}-T_{k}).\nonumber
\end{eqnarray}
Hence $B_n=1+\frac{1}{2}(T_n-T_{n-1})$ is also correct for
$n=k+1$, which completes the induction.

Now one can calculate the following conditional expectation
\begin{eqnarray}
\mbox{\boldmath
$E$}\left[T_{n+1}-T_{n}+2|T_{n}-T_{n-1}+2\right]&=&
T_{n}-T_{n-1}+2+\mbox{\boldmath
$E$}\left[\left.\sum_{j=1}^{B_n}\tau_j^{(n+1)}\right|
T_{n}-T_{n-1}+2\right]\nonumber\\
&=&(1+{\textstyle\frac{1}{2}}\mbox{\boldmath
$E$}T_1)(T_{n}-T_{n-1}+2).\nonumber
\end{eqnarray}
Hence the sequence
\[
Y_n=\frac{T_n-T_{n-1}+2}{(1+\frac{1}{2}\mbox{\boldmath
$E$}T_1)^n},\quad n=1,2,3,\ldots,
\]
is a positive martingale.

Since $\mbox{\boldmath $E$}Y_n=\mbox{\boldmath $E$}Y_1=2$, we have
$\Pr[Y_n<\infty]=1$, for any $n$, which implies the almost sure
finiteness of the return times $T_n$, $n=1,2,\ldots$.

The relationship $\sup_{n}\mbox{\boldmath $E$}(Y_n^2)<\infty$ is a
well-known sufficient condition for the uniform integrability of
the sequence $\{Y_n\}_{n=1}^\infty$, see, e.g., the book by
Shiryaev \cite{s98}. In our case Eq.\ (\ref{tn}) yields
\begin{eqnarray}
s_{n+1}&\equiv&\mbox{\boldmath $E$}(T_{n+1}-T_{n}+2)^2\nonumber\\
&=&\mbox{\boldmath $E$}(T_{n}-T_{n-1}+2)^2+2\mbox{\boldmath
$E$}\left[(T_{n}-T_{n-1}+2)
\sum_{j=1}^{B_n}\tau_j^{(n+1)}\right]+\mbox{\boldmath $E$}
\left(\sum_{j=1}^{B_n}\tau_j^{(n+1)}\right)^2.\nonumber
\end{eqnarray}
On calculating the expected values with the help of the tower
property we obtain the following simple recurrent relationship
\[
s_{n+1}=s_n(1+{\textstyle\frac{1}{2}}\mbox{\boldmath
$E$}T_1)^2+\mbox{Var}(T_1)(1+{\textstyle\frac{1}{2}}\mbox{\boldmath
$E$}T_1)^n.
\]
Solving the recurrent relationship we obtain
\begin{equation}
s_{n+1}=2\left(2+\frac{\mbox{\boldmath $E$}T_1^2}{\mbox{\boldmath
$E$}T_1}\right)(1+{\textstyle\frac{1}{2}}\mbox{\boldmath
$E$}T_1)^{2n+1}- 2\frac{\mbox{Var}(T_1)}{\mbox{\boldmath $E$}T_1}
\left(1+{\textstyle\frac{1}{2}}\mbox{\boldmath $E$}T_1\right)^n.
\label{ey2}
\end{equation}
Hence
\[
\sup_{n}\frac{s_n}{(1+\frac{1}{2}\mbox{\boldmath
$E$}T_1)^{2n}}<\infty,
\]
implying the uniform integrability of the martingale $Y_n$,
$n=1,2,\ldots$.

\hfill \rule{2mm}{2mm}\newline

{\sf Corollary 1.} Let $q+2p<1$, then for almost all initial
arrangements of the arrows
\[
\lim_{n\to\infty}\frac{T_n-T_{n-1}+2}{(1+\frac{1}{2}\mbox{\boldmath
$E$}T_1)^n}=Y,
\]
where $Y$ is a random variable with a proper distribution
($\Pr[Y<\infty]=1$). The expected value and the variance of the
random variable $Y$ are given by
\[
\mbox{\boldmath $E$}Y=2,\quad\mbox{Var}(Y)=
\frac{4\mbox{Var}(T_1)}{\mbox{\boldmath $E$}T_1 (2+\mbox{\boldmath
$E$}T_1)}.
\]
{\sf Proof.} Since the sequence $\{Y_n\}_{n=1}^\infty$ is a
positive martingale, the Doob martingale convergence theorem, see,
e.g.,\ the book by Shiryaev \cite{s98}, tells us that
$\lim_{n\to\infty}Y_n=Y$, where $Y$ is a random variable with a
proper distribution. Since the sequence $\{Y_n\}_{n=1}^\infty$ is
uniformly integrable $\mbox{\boldmath
$E$}Y=\lim_{n\to\infty}\mbox{\boldmath $E$}Y_n=2$.

Equation (\ref{ey2}) yields
\[
\lim_{n\to\infty}\mbox{\boldmath
$E$}Y_n^2=4+\frac{4\mbox{Var}(T_1)}{\mbox{\boldmath
$E$}T_1(2+\mbox{\boldmath $E$}T_1)}.
\]
To show that $\mbox{\boldmath
$E$}Y^2=\lim_{n\to\infty}\mbox{\boldmath $E$}Y_n^2$, we need the
uniform integrability of the sequence $\{Y_n^2\}_{n=1}^\infty$. To
that end one can use the sufficient condition
$\sup_{n}\mbox{\boldmath $E$}Y_n^3=\sup_{n}\mbox{\boldmath
$E$}(Y_n^2)^{3/2}<\infty$. One can check by a direct calculation
similar to that used in the proof of Theorem 1 that the sufficient
condition is indeed satisfied. Hence
\[
\mbox{Var}(Y)=\mbox{\boldmath $E$}Y^2-(\mbox{\boldmath $E$}Y)^2=
\frac{4\mbox{Var}(T_1)}{\mbox{\boldmath $E$}T_1 (2+\mbox{\boldmath
$E$}T_1)}.
\]
\hfill\rule{2mm}{2mm}\newline

{\sf Corollary 2.} Let $q+2p<1$, then
\begin{eqnarray}
&&\mbox{\boldmath $E$}T_n =2\,\frac{2+\mbox{\boldmath
$E$}T_1}{\mbox{\boldmath
$E$}T_1}\left[\left(1+{\textstyle\frac{1}{2}}\mbox{\boldmath
$E$}T_1\right)^n-1\right]-2n,\nonumber\\
&&\mbox{Var}(T_n) \sim
\frac{\mbox{Var}(T_1)}{\left(\frac{1}{2}\mbox{\boldmath
$E$}T_1\right)^3} \left(1+{\textstyle\frac{1}{2}}\mbox{\boldmath
$E$}T_1\right)^{2n+1}, \label{vartn}\\
&&\lim_{n\to\infty}
\frac{T_n}{\left(1+{\textstyle\frac{1}{2}}\mbox{\boldmath
$E$}T_1\right)^{n+1}}=\frac{2Y}{\mbox{\boldmath
$E$}T_1},\quad\mbox{almost surely,}\nonumber
\end{eqnarray}
where the random variable $Y$ is identical to the one from
Corollary 1.

{\sf Proof.} Recall that the martingale $\{Y_k\}_{k=1}^\infty$ is
defined by
\[
Y_k=\frac{T_k-T_{k-1}+2}{(1+\frac{1}{2}\mbox{\boldmath
$E$}T_1)^k}.
\]
Taking the denominator to the l.h.s.\ and summing over $k$ from
$1$ to $n$ one obtains
\begin{equation}
T_n=\sum_{k=1}^{n}\left(1+{\textstyle\frac{1}{2}}\mbox{\boldmath
$E$}T_1\right)^kY_k-2n,
\label{tn2}
\end{equation}
where we have used $T_0=0$. Since $\mbox{\boldmath $E$}Y_l=2$, a
summation of the geometric series yields
\[
\mbox{\boldmath $E$}T_n=2\,\frac{2+\mbox{\boldmath
$E$}T_1}{\mbox{\boldmath
$E$}T_1}\left[\left(1+{\textstyle\frac{1}{2}}\mbox{\boldmath
$E$}T_1\right)^n-1\right]-2n.
\]

Equation (\ref{tn2}), the martingale property $\mbox{\boldmath
$E$}(Y_k|Y_f)=Y_f$, for $f<k$, and straightforward calculations
yield the main asymptotics of the variance $\mbox{Var}(T_n)$, Eq.\
(\ref{vartn}).

Since $\lim_{k\to\infty}Y_k=Y$ (almost surely), an application of
a standard technique from analysis to Eq.\ (\ref{tn2}) yields
\[
\lim_{n\to\infty}
\frac{T_n}{\left(1+{\textstyle\frac{1}{2}}\mbox{\boldmath
$E$}T_1\right)^{n+1}}= \lim_{n\to\infty}
\sum_{k=1}^{n}\left(1+{\textstyle\frac{1}{2}}\mbox{\boldmath
$E$}T_1\right)^{k-n-1}Y_k =\frac{2Y}{\mbox{\boldmath $E$}T_1}.
\]
\hfill \rule{2mm}{2mm}\newline


The last Corollary describes in detail the large-$n$ behaviour of
the $n^{\rm th}$ return time $T_n$ in the subcritical regime,
where $\mbox{\boldmath $E$}T_1<\infty$. The following crude bound
will be helpful at the critical point.

{\sf Corollary 3.}
\begin{equation}
T_n-T_{n-1}\leq T_n \leq 2(T_n-T_{n-1}).
\label{tntn}
\end{equation}

{\sf Proof.} For the number of buds on the $n^{\rm th}$ return (to
the root) path we have $B_n\geq 2B_{n-1}$, hence $B_{n-l}\leq
2^{-l}B_n$. Summing Eq.\ (\ref{bntn}) we obtain
\[
T_n=2\left(\sum_{l=1}^n B_l-n\right).
\]
Therefore $T_n\leq 4B_n-2n$, and using Eq.\ (\ref{bntn}) again we
obtain
\[
T_n-T_{n-1}\leq T_n\leq 2(T_n-T_{n-1}).
\]
\hfill \rule{2mm}{2mm}\newline

Theorem 1 and its corollaries give a fairly comprehensive
description of the frequency of return to the root. Our next aim
is a description of the height of the domain of visited sites.
Remark 1 describes the distribution of the highest visited
generation at time $T_1$. Investigation of the height of the
domain at later times is a much more delicate problem.
We will find the asymptotic behaviour of the density of visited
sites, $v_k(t)$, in the $k^{\rm th}$ generation of the tree,
defined as the ratio of the number of sites visited by time $t$ to
the total number of sites, $2^{k-1}$, in the $k^{\rm th}$
generation.

In order to describe the growth of the domain of visited sites on
the Cayley tree, let us consider an arbitrary branch
$w_n=(e_1,e_2,\ldots,e_n)$ of the tree, where $e_l$,
$l=1,2,\ldots,n$ are the segments (edges) of the branch, see Fig.\
5. With any edge $e_l$ one can associate an ``energy"
$\varepsilon_l$ as follows. The energy of a left edge $e_l$ (like
$e_a$ on Fig.\ 5) is equal to 0, if the arrow at the bottom of the
edge $e_l$ points along the edge, and $\varepsilon_k=1$ otherwise.
The energy of a right edge $e_l$ (like $e_b$ on Fig.\ 5) is equal
to 1, if the arrow at the bottom of the edge $e_l$ points down,
and $\varepsilon_l=0$ otherwise. In other words, the energy of an
edge $e_l$ is equal to 1, if the arrow at the bottom of the edge
causes the walker to deviate from the Euler circuit, and the
energy is equal to 0 if the walker passes the edge
``effortlessly". The energy of a branch $w_n$ is the sum of the
energies of its edges.

\begin{figure}[t]
\setlength{\unitlength}{1mm}
\begin{picture}(150,75)(0,-70)

\multiput(77.2,-54)(-1.46,0.73){22}{\makebox(0,0){\circle*{0.2}}}
\multiput(45.2,-38)(1,1){16}{\makebox(0,0){\circle*{0.2}}}
\multiput(45.2,-38)(-1,1){16}{\makebox(0,0){\circle*{0.2}}}
\multiput(109.2,-38)(-1,1){16}{\makebox(0,0){\circle*{0.2}}}

\multiput(29.2,-22)(1,1.5){8}{\makebox(0,0){\circle*{0.2}}}
\multiput(29.2,-22)(-1,1.5){8}{\makebox(0,0){\circle*{0.2}}}
\multiput(61.2,-22)(1,1.5){8}{\makebox(0,0){\circle*{0.2}}}
\multiput(61.2,-22)(-1,1.5){8}{\makebox(0,0){\circle*{0.2}}}
\multiput(93.2,-22)(-1,1.5){8}{\makebox(0,0){\circle*{0.2}}}
\multiput(93.2,-22)(1,1.5){8}{\makebox(0,0){\circle*{0.2}}}
\multiput(125.2,-22)(1,1.5){8}{\makebox(0,0){\circle*{0.2}}}

\multiput(21.2,-10)(0.72,1.44){7}{\makebox(0,0){\circle*{0.2}}}
\multiput(21.2,-10)(-0.72,1.44){7}{\makebox(0,0){\circle*{0.2}}}
\multiput(37.2,-10)(0.72,1.44){7}{\makebox(0,0){\circle*{0.2}}}
\multiput(37.2,-10)(-0.72,1.44){7}{\makebox(0,0){\circle*{0.2}}}
\multiput(53.2,-10)(0.72,1.44){7}{\makebox(0,0){\circle*{0.2}}}
\multiput(53.2,-10)(-0.72,1.44){7}{\makebox(0,0){\circle*{0.2}}}
\multiput(69.2,-10)(0.72,1.44){7}{\makebox(0,0){\circle*{0.2}}}
\multiput(69.2,-10)(-0.72,1.44){7}{\makebox(0,0){\circle*{0.2}}}
\multiput(85.2,-10)(0.72,1.44){7}{\makebox(0,0){\circle*{0.2}}}
\multiput(85.2,-10)(-0.72,1.44){7}{\makebox(0,0){\circle*{0.2}}}
\multiput(101.2,-10)(0.72,1.44){7}{\makebox(0,0){\circle*{0.2}}}
\multiput(117.2,-10)(0.72,1.44){7}{\makebox(0,0){\circle*{0.2}}}
\multiput(101.2,-10)(-0.72,1.44){7}{\makebox(0,0){\circle*{0.2}}}
\multiput(133.2,-10)(0.72,1.44){7}{\makebox(0,0){\circle*{0.2}}}
\multiput(133.2,-10)(-0.72,1.44){7}{\makebox(0,0){\circle*{0.2}}}

\multiput(16,0)(16,0){8}{{\circle*{0.2}}}
\multiput(26,0)(16,0){8}{{\circle*{0.2}}}

\put(77,-70){\circle{1}} \put(70,-62){$e_1$}
\put(80,-62){$\varepsilon_1=0$}

\put(87,-44){$e_2$} \put(95,-49){$\varepsilon_2=0$}

\put(112,-28){$e_3$} \put(120,-33){$\varepsilon_3=0$}

\put(20,-17){$\varepsilon_a$} \put(35.2,-17){$\varepsilon_b$}

\put(122,-14){$e_4$} \put(107,-18){$\varepsilon_4=1$}

\put(109.5,-6){$e_5$} \put(108,3){$\varepsilon_5=1$}

\put(77,-54){{\circle{1}}}
\multiput(45,-38)(64,0){2}{{\circle{1}}}
\multiput(29,-22)(32,0){4}{{\circle{1}}}
\multiput(21,-10)(16,0){8}{{\circle{1}}}
\multiput(16,0)(16,0){8}{{\circle{1}}}
\multiput(26,0)(16,0){8}{{\circle{1}}}

\put(77,-70){\vector(0,1){4.6}} \put(77,-54){\vector(2,1){4.2}}
\put(109,-38){\vector(-1,1){3.6}}
\put(125,-22){\vector(-1,-1){3.6}}
\put(117,-10){\vector(1,2){1.8}} \put(112,0){\vector(1,-2){1.8}}

\thicklines

\put(77,-70){\line(0,1){16}} \put(77,-54){\line(2,1){32}}
\put(109,-38){\line(1,1){16}} \put(125,-22){\line(-2,3){8}}
\put(117,-10){\line(-1,2){5}}

\put(29,-22){\line(2,3){8}} \put(29,-22){\line(-2,3){8}}
\end{picture}
\caption{A branch $w_5$ (path) of the Cayley tree, its edges
$(e_1,e_2,\ldots,e_5)$, and the associated random ``energies"
$(\varepsilon_1,\varepsilon_2,\ldots,\varepsilon_5)$. The energies
of edges growing from the same site of the tree, like
$\varepsilon_a$ and $\varepsilon_b$, are not independent.}
\end{figure}
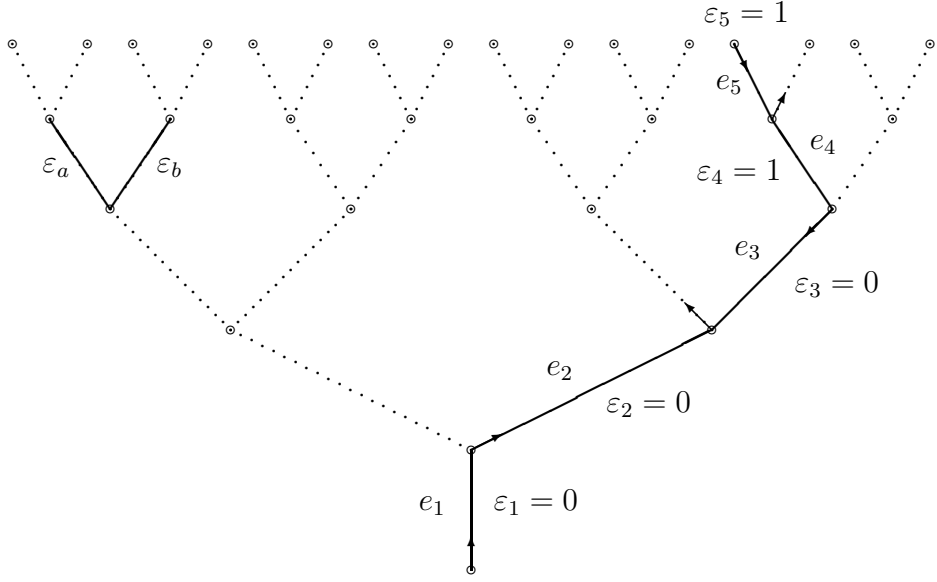

The domain of visited sites swallows up the edges of a path $w_n$ as
follows. During the time interval $[0,T_1]$ (before the first
return to the root) the domain swallows all the edges of the path
$w_n$ till the first obstacle --- the first edge $e_l$ with
$\varepsilon_l=1$. During the time interval $[T_1,T_2]$ (after the
first return but before the second return to the root) the domain
of visited sites swallows up the edge $e_l$ and all zero-energy edges
which follow $e_l$ until the second obstacle --- the second edge
$e_m$ with non-zero energy, and so on. During the time interval
$[T_j,T_{j+1}]$ (after the $j^{\rm th}$ return but before the
$j+1^{\rm th}$ return to the root) the domain of visited sites
swallows up all the edges between the $j^{\rm th}$ and $j+1^{\rm th}$
edges with non-zero energy. Thus, the number of visited sites in
the $k^{\rm th}$ generation at time $T_m$ is equal to the number
of paths $w_k$ with less than $m$ obstacles, or, equivalently,
with the path energies $E(w_k)=\sum_{l:e_l\in w_k}\varepsilon_l$
less than $m$.

Let us consider the following sum (partition function)
\[
\Theta_k=\sum_{w_k}\exp[-\beta E(w_k)],
\]
where the summation runs over all branches $w_k$ of a tree with
$k$ generations. We have
\[
\Theta_k=\sum_{n=0}^k\#\{w_k:E(w_k)=n\}\exp[-\beta n].
\]
Hence, the large $k$ limit of $k^{-1}\ln\Theta_k$ is the
Legendre-transform of
\[
\nu(y)\equiv\lim_{k\to\infty}k^{-1}\ln\#\{w_k:E(w_k)=[ky]\},
\]
where $[ky]$ is the integer part of $ky$.

On the other hand, the sum $\Theta_n$ is almost identical to the
partition function of a directed polymer on a Cayley tree, see
\cite{b93}. The difference between $\Theta_n$ and the partition
function in \cite{b93} is that not all the energies
$\varepsilon_l$ are independent. Indeed if two edges $e_a$ and
$e_b$ grow from the same site of the tree, see Fig.\ 5, then
\begin{eqnarray}
&&\Pr[\varepsilon_a=1,\varepsilon_b=1]=1-p-q,\quad
\Pr[\varepsilon_a=0,\varepsilon_b=1]=0,\nonumber\\
&&\Pr[\varepsilon_a=1,\varepsilon_b=0]=q,
\quad\mbox{and}\quad\Pr[\varepsilon_a=0,\varepsilon_b=0]=p.\nonumber
\end{eqnarray}
Nevertheless, the large-$k$ asymptotics of $k^{-1}\ln\Theta_k$ can
be found by virtually verbatim repetition of the derivation from
\cite{b93}. In particular, if we denote ${\cal A}_k$ the
$\sigma$-algebra generated by the random energies of the first $k$
generations of the tree, and define
\[
M_k=\frac{\Theta_k}{\left[(2-2p-q)e^{-\beta}+2p+q\right]^{k-1}},
\]
then the stochastic sequence $\{M_k,{\cal A}_k\}_{k=1}^\infty$ is
a positive martingale, and $\mbox{\boldmath $E$}M_k=1$.

Using the martingale technique from \cite{b93} we obtain.
\vspace{\abovedisplayskip}

{\sf Proposition 1.} If $0\leq 2p+q< 1$, then
\begin{equation}
f(\beta)\equiv\lim_{k\to\infty}k^{-1}\ln\Theta_k= \left\{
\begin{array}{cl}
\ln\left[(2-2p-q)e^{-\beta}+2p+q\right],&\mbox{ if
}\beta\leq\beta_c;\vspace{1mm}\\
\frac{\beta}{\beta_c}\ln\left[(2-2p-q)e^{-\beta_c}+2p+q\right],&\mbox{
if }\beta\geq\beta_c;
\end{array}
\right. \label{fe1}
\end{equation}
where $\beta_c$ is the positive solution of
\[
\ln\left[(2-2p-q)e^{-\beta}+2p+q\right]=
\frac{\beta(2p+q)e^{\beta}}{2-2p-q+(2p+q)e^{\beta}}.
\]
While if $1\leq 2p+q\leq 2$, then
\begin{equation}
f(\beta)\equiv\lim_{k\to\infty}k^{-1}\ln\Theta_k=\ln\left[(2-2p-q)e^{-\beta}+2p+q\right].
\label{fe2}
\end{equation}
\hfill \rule{2mm}{2mm}

{\sf Lemma 3.} The logarithmic asymptotics of the number of path
$w_k$ with the energy $[ky]$, $y\in(0,1)$ is given by
\begin{eqnarray}
\nu(y)&\equiv&\lim_{k\to\infty}k^{-1}\ln\#\{w_k:E(w_k)=[ky]\}\nonumber\\
&=& \left[y\ln\frac{2-(2p+q)}{y}
+(1-y)\ln\frac{2p+q}{1-y}\right]^+, \label{nuy}
\end{eqnarray}
where $[x]^+=\max(x,0)$ is the positive part of $x$.

{\sf Proof.} The free energy $f(\beta)$, given by Eqs.\
(\ref{fe1}) and (\ref{fe2}), is the Legendre transform of the
logarithmic asymptotics $\nu(y)$. Namely
\[
f(\beta)=\max_{y\in[0,1]}\left[-\beta y+\nu(y)\right].
\]
Therefore
\[
\nu(y)=\min_{\beta\geq0}\left[\beta y+f(\beta)\right].
\]
Solving the minimization problem we obtain Eq.\ (\ref{nuy}).

\hfill \rule{2mm}{2mm}

An inspection of the function $\nu(y)$ shows that there are around
$[2-(2p+q)]^k$ branches $w_k$ containing $k$ obstacles for the
walker to overcome. At the same time there are a few branches with
only around $[y^*k]$ obstacles, where $y^*\in(0,1)$ is a solution
of the equation
\[
y\ln\frac{2-(2p+q)}{y}+(1-y)\ln\frac{2p+q}{1-y}=0.
\]
Hence, there exists a growing with time gap, of the width
$m(1/y^*-1)$ generations at time $T_m$, between the highest
visited generation and the highest completely explored generation
of the Cayley tree. Therefore, neither generation is likely to be
a sensible measure of the height of the domain of visited sites.

It is a common practice in situations like that to concentrate
ones attention on typical branches of the tree. Therefore, we
define the height of the domain of visited sites as a number
(function) $H(t)\sim h\ln t$, such that the density of visited
sites in generation $x\ln t$ at time $t$, $v_{x\ln t}(t)$, tends
to zero with $t$ if $x>h$, and $v_{x\ln t}(t)\to 1$, if $x<h$. We
will see shortly that this definition is a sensible one for the
problem under consideration. Of course, the choice of the
asymptotic form $H(t)\sim h\ln t$ is specific to Cayley trees, and
was actually made after the density of visited sites was
calculated.

The logarithmic asymptotics $\nu(y)$ attains its maximum, $\ln 2$,
at $y=p+\frac{1}{2}q$. Hence, the typical branches $w_k$ have the
energy $E(w_k)\sim k(p+\frac{1}{2}q)$. Thus, the domain of visited
sites swallows up a typical branch $w_k$ of the tree after
$k(p+\frac{1}{2}q)$ returns to the root.

{\sf Theorem 2.} Let $q+2p<1$, then the height of the domain of
visited sites, $H(t)$, grows as logarithm of time,
\[
H(t)\sim\frac{\ln t}{(p+\frac{1}{2}q)\ln\left(1+\frac{1}{2}
\mbox{\boldmath $E$}T_1\right)}.
\]
{\sf Proof.} As follows from Corollary 2, the number of returns to
the root by time $t$ for the walker is given by
\[
m\sim \frac{\ln t}{\ln\left(1+\frac{1}{2}\mbox{\boldmath
$E$}T_1\right)},
\]
as $t\to\infty$. The asymptotic number of obstacles in a typical
branch $w_k$ of the Cayley tree is given by $k(p+\frac{1}{2}q)$,
as $k\to\infty$. Hence, the typical penetration after $m$ returns
to the root is approximately $m/(p+\frac{1}{2}q)$ generations,
while the typical penetration by time $t$ is
\[
H(t)\sim \frac{\ln
t}{(p+\frac{1}{2}q)\ln\left(1+\frac{1}{2}\mbox{\boldmath
$E$}T_1\right)}\quad\mbox{generations.}
\]
\hfill \rule{2mm}{2mm}

Unfortunately it is difficult to go beyond the logarithmic
asymptotics $\nu(y)$ of 
the number of paths $w_k$ with the energy $E(w_k)=[ky]$.
Nevertheless, one can guess that the number of paths with the
energy $E(w_k)\sim k(p+\frac{1}{2}q)+\sqrt{k}u$ is controlled
entirely by the quadratic term in the Taylor expansion for
$\nu(y)$ at $y=p+\frac{1}{2}q$. If this is indeed the case then,
in the spirit of the local limit theorem, we obtain
\[
\#\left\{w_k:E(w_k)=k(p+{\textstyle\frac{1}{2}}q)+
\sqrt{k}u\right\}\sim
\frac{c} {\sqrt{k}} \exp\left[k\nu(p+{\textstyle\frac{1}{2}}q)+
{\textstyle\frac{1}{2}}\nu''(p+{\textstyle\frac{1}{2}}q)u^2\right]
\]
\begin{equation}
=\frac{2^kc}{\sqrt{k}} \exp\left[-
\frac{u^2}{2(p+{\textstyle\frac{1}{2}}q)
(1-p-{\textstyle\frac{1}{2}}q)}\right].
\label{llt}
\end{equation}

The density of visited sites in generation $n$ at time $T_m$ is
given by
\[
v_n(T_m)=\frac{1}{2^{n-1}}\sum_{f<m}\#\left\{w_n:E(w_n)=f\right\}.
\]
Approximating the sum by an integral (very much like in the normal
approximation to the binomial distribution) and taking into
account Eq.\ (\ref{llt}) one obtains
\begin{equation}
v_n(T_m)\sim\frac{1}{\sqrt{2\pi\sigma^2}}
\int_{-\infty}^{[m-n(p+q/2)]/\sqrt{n}}dx\,\exp\left(-\frac{x^2}{2\sigma^2}\right),
\label{dvs}
\end{equation}
where $\sigma^2=(p+{\textstyle\frac{1}{2}}q)
(1-p-{\textstyle\frac{1}{2}}q)$.

We summarize the above discussion by a hypothesis which might well
be true. \vspace{\abovedisplayskip}

{\sf Hypothesis 1.} The width of the boundary of the domain of
visited sites of size $n$ generations grows with $n$ as
$\sqrt{n}$. The drop of the density of visited sites on the
boundary from 1 to 0 is described by the error function, see Eq.\
(\ref{dvs}).

\hfill \rule{2mm}{2mm}

Note that at the critical point $2p+q=1$ the variance $\sigma^2$
in Eq.\ (\ref{dvs}) reaches its maximal value, $\frac{1}{4}$, but
remains finite. Therefore the density profile of the domain of
visited sites does not disintegrate as we approach the critical
point. Instead, as $2p+q$ approaches 1, the walker tends to spend
more and more time in long (low-density) excursions away from the
compact domain of visited sites. Those long excursions do not
create new compact visited domains, somewhat like water poured
into sand does not create puddles.

\section{The low-density phase.}

Let now $q+2p>1$. In this case the associated branching process
degenerates with probability $x^*$ which is a solution of the
equation $x=g(x)$ less than 1, see \cite{h91,f70}, where the
function $g(x)$ is given by Eq.\ (\ref{gfg}). That is,
$x^*=(1-q-p)/p$. A routine application of the Borel-Cantelli lemma
shows that in this case, with probability 1, the Euler walker
visits the root (and any given generation of the tree) only a
finite number of times.

Let $k$ be large enough to guarantee that only one copy of the
associated branching process --- the copy which does not
degenerate --- has survived until the $k^{\rm th}$ generation.
Then the number of visited sites in the $k^{\rm th}$ generation,
$V_k$, (after the last visit of the $k^{\rm th}$ generation) does
not exceeds the number of particles in a single copy of the
associated branching process. Namely, $V_k\leq W(q+2p)^k$, where
$W$ is a random variable with a proper distribution
($P[W<\infty]=1$). Since $q+2p<2$ unless $p=1$, we have
$V_k/2^k\to0$ as $k\to\infty$. That is, the model is in the
low-density phase when $q+2p>1$.

The bound $V_k\leq W(q+2p)^k$ is a gross overestimation of the
number of visited sites. Most likely $V_k$ does not grow faster
than something like a constant times $\ln k$.

\section{The critical point.}

In this section we consider the critical case $q+2p=1$. Like in
the subcritical case $q+2p<1$, the associated branching process
degenerates with probability 1 if $q+2p=1$. However the branching
process becomes critical, and its properties differ substantially
from those in the subcritical regime. As we shall see shortly, the
first moments of all relevant random variables are infinite if
$q+2p=1$. As a consequence, extraction of properties of the random
variables from their generating functions is no longer
straightforward. \vspace{\abovedisplayskip}

{\sf Lemma 4.} Let $q+2p=1$, then the walker returns to the root
for the first time at a finite (almost surely) time-instant $T_1$,
such that
\begin{equation}
\Pr[T_1=2k]\sim \frac{1}{2\sqrt{\pi p}}\,k^{-3/2},\quad\mbox{as
}k\to\infty.
\label{at1}
\end{equation}
{\sf Proof.} Analogously to the subcritical case,
the probability generating function of the total number of
descendants, $Z$, is given by
\begin{equation}
f(x)=1+\frac{1}{2px}\left[1-x-\sqrt{(1-x)[1-(1-4p)x]}\right].
\label{fcx}
\end{equation}
Using Eq.\ (\ref{pas}) we obtain
\[
\Pr[T_1=2k]=\Pr[Z=k]\sim \frac{1}{2\sqrt{\pi
p}}\,k^{-3/2},\quad\mbox{as }k\to\infty.
\]
\hfill \rule{2mm}{2mm}\newline

{\sf Remark 2.} The large-$k$ asymptotics of $\Pr[T_1=2k]$ makes
it clear that $\mbox{\boldmath $E$}T_1=\infty$. It is still
desirable to have a deterministic measure indicating likely values
of the first-return time $T_1$. For that purpose one can use the
quantiles $Q_1(x)$ --- solutions of the equation $\Pr[T_1\leq
Q_1]=x$. The asymptotic formula (\ref{at1}) yields the following
equation for approximate values of $Q_1(x)$
\[
\frac{1}{2\sqrt{\pi p}}\sum_{k>Q_1/2}^\infty k^{-3/2}=1-x.
\]
Replacing the sum by an integral and solving the obtained equation
for $Q_1(x)$ one obtains $Q_1(x)\approx \frac{2}{\pi p (1-x)^2}$.
For values of $x$ close to 1, the precision of the found
approximation for $Q_1(x)$ is quite reasonable. For instance, in
the case $p=0.1$ it gives $Q_1(\frac{3}{4})\approx 102$, while the
exact value is $Q_1(\frac{3}{4})=98$.

\hfill \rule{2mm}{2mm}

{\sf Remark 3.} Like in the subcritical case, see Remark 1,
asymptotic properties of the distribution of the height of the
first-return path, $H_1$, follow from standard results of the
theory of branching processes. It is shown in the book by Harris
\cite{h91} that the large-$k$ asymptotics of $\Pr[Z_k=0]$ in the
case $\mbox{\boldmath $E$}X=1$ is given by
\[
1-\Pr[Z_k=0]\sim \frac{1}{pk}.
\]
Hence, the distribution of $H_1$ displays a power-law decay,
\[
\Pr[H_1=k]=\Pr[Z_{k+1}=0]-\Pr[Z_k=0]\sim\frac{1}{pk^2}.
\]
\hfill \rule{2mm}{2mm}

In order to investigate the distribution of the return to the root
instants $T_2,T_3,\ldots$ let us first find the probability
generating functions $G_2(x),G_3(x),\ldots$ for the number of buds
$B_2,B_3,\ldots$ on the corresponding paths. Using Eq.\ (\ref{bn})
and the tower property one obtains
\[
G_n(x)=\mbox{\boldmath $E$}x^{B_n}=\mbox{\boldmath
$E$}(xf(x))^{B_{n-1}}=G_{n-1}(\varphi(x)),
\]
where $\varphi(x)\equiv xf(x)$ is the generating function of
$B_1$, and $f(x)$ is given by Eq.\ (\ref{fcx}). It is clear now
that $G_n(x)$ is the $n^{\rm th}$ iteration of $\varphi(x)$, that
is,
\[
G_n(x)=\underbrace{\varphi(\varphi(\ldots\varphi(x)\ldots))}_{
\mbox{$n$ times}}.
\]
Hence $G_n(x)=\varphi(G_{n-1}(x))$ as well.

\vspace{\abovedisplayskip}

{\sf Theorem 3.} Let $q+2p=1$, then the walker returns to the root
infinitely often at (almost surely) finite time instants
$T_1,T_2,T_3,\ldots$. Moreover
\begin{equation}
\Pr[T_n-T_{n-1}=2k]\sim
\frac{1}{2^n\Gamma(1-2^{-n})p^{1-2^{-n}}k^{1+2^{-n}}}
\quad\mbox{as }k\to\infty. \label{exps}
\end{equation}

{\sf Proof.} The probability $\Pr[T_n-T_{n-1}=2k]$ is given by the
integral
\[
\Pr[T_n-T_{n-1}=2k]=\frac{1}{2\pi
i}\int_C\!\frac{G_n(z)}{z^{k+2}}\,dz,
\]
where $C$ is a sufficiently small closed contour encircling $0$,
and $G_n(x)$ is the probability generating function of $B_n$. To
find the large-$k$ asymptotics of this integral we adapt the
contour integration from \cite{o49}. For that purpose we have to
know analytical properties of the generating functions $G_n(x)$.

By definition
\[
G_n(z)=\sum_{l=0}^\infty \Pr[B_n=l]z^l,
\]
hence the function $G_n(z)$ is analytic inside the unit circle
$\{z:|z|<1\}$. Since $G_{n-1}(1)=1$, and
$G_{n}(z)=\varphi(G_{n-1}(z))$, the point $z=1$ is a branch point
of $G_n(z)$. Since $|G_n(e^{ix})|<1$ for any real $x\in (0,2\pi
)$, the point $z=1$ is the only singularity of the function
$G_n(z)$ on the boundary of the unit circle $\{z:|z|<1\}$.

From the explicit formula for the function $\varphi(z)$ it is
clear that the generating function $G_n(z)$ has only a finite
number of points of non-analyticity. Hence, there exists a disc
$A_n=\{z:|z|\leq \alpha_n\}$, with $\alpha_n>1$, such that $z=1$
is the only singularity of the functions $G_k(z)$,
$k=1,2,\ldots,n$ in $A_n$.

Denote $D_n$ the boundary of the disc $A_n$ with a radial cut
running outwards from $x=1$. The generating function $G_n(z)$ can
be written as follows
\begin{equation}
G_n(z)=1-a_n(1-z)^{2^{-n}}+(1-z)^{2^{-n+1}}f_n(z), \label{gfr}
\end{equation}
where $f_n(z)$ is analytic and bounded inside $D_n$: $|f_n(z)|\leq
b(p)<\infty$. Indeed, we already know that the function $G_n(z)$
is analytic inside $D_n$. Since
\[
f_n(z)=\frac{G_n(z)-1+a_n(1-z)^{2^{-n}}}{(1-z)^{2^{-n+1}}},
\]
it must be analytic inside $D_n$ as well.

To show that $f_n(z)$ is bounded inside $D_n$ we can use
induction. The function $f_1(z)$ is obviously bounded in any
circle with finite radius. Assume now that $f_n(z)$ is bounded in
any circle with finite radius for $n=k$, then for $n=k+1$ we
obtain
\[
G_{k+1}(z)=\varphi(G_{k}(z))=1-\sqrt{\frac{a_k}{p}}(1-z)^{2^{-k-1}}
+(1-z)^{2^{-k}}f_{k+1}(z),
\]
where
\[
f_{k+1}(z)=\left(\frac{1}{2p}-1\right)
\left[a_k-(1-z)^{2^{-k}}f_k(z)\right]
\]
\[
-(1-z)^{-2^{-k-1}}\left[\frac{1}{2p}\sqrt{\left[a_k-(1-z)^{2^{-k}}f_k(z)\right]
[1-(1-4p)G_k(z)]}-\sqrt{\frac{a_k}{p}}\right].
\]
Hence $f_{k+1}(z)$ is bounded in any circle with finite radius as
well, completing the induction.

From the above equations we obtain the recurrent relationship
$a_{k+1}=\sqrt{a_k p^{-1}}$, with the initial condition
$a_1=\sqrt{p^{-1}}$. The solution of this recurrent relationship
is given by $a_k=p^{-1+2^{-k}}$.

On substitution of Eq.\ (\ref{gfr}) in the integral representation
for the probability $\Pr[T_n-T_{n-1}=2k]$ we obtain
\[
\Pr[T_n-T_{n-1}=2k]=-\frac{a_n}{2\pi i}\int_C
\frac{(1-z)^{2^{-n}}}{z^{k+2}}\,dz+ \frac{1}{2\pi
i}\int_C\frac{(1-z)^{2^{-n+1}}f_n(z)}{z^{k+2}}\,dz=
\]
\[
=(-1)^{k}a_n\left(\!\!\begin{array}{c}2^{-n}\\k+1\end{array}\!\!\right)
+ \frac{1}{2\pi
i}\int_{D_n}\!\frac{(1-z)^{2^{-n+1}}f_n(z)}{z^{k+2}}\,dz.
\]
Since the function $f_n(z)$ is bounded inside $D_n$, the remaining
integral is of the same order as
\[
\int_1^{\alpha_n}\!\frac{(1-x)^{2^{-n+1}}}{x^{k+2}}\,dx=
O\left(\!\!\begin{array}{c}2^{-n+1}\\k+1\end{array}\!\!\right).
\]
Therefore
\[
\Pr[T_n-T_{n-1}=2k]\sim
\frac{1}{2^n\Gamma(1-2^{-n})p^{1-2^{-n}}k^{1+2^{-n}}}
\quad\mbox{as }k\to\infty.
\]

Finally, note that $\Pr[T_n-T_{n-1}<\infty]=1$, and according to
Eq.\ (\ref{tntn}) we have $T_n\leq 2(T_n-T_{n-1})$. Hence, all
return to the root instants $T_n$ are almost surely finite.

\hfill \rule{2mm}{2mm}

{\sf Theorem 4.} Let $q+2p=1$, then the median of the height of
the domain of visited sites grows with time as the iterated
logarithm $2\log_2\log_2 t$.

{\sf Proof.} If $q+2p=1$, then the number of obstacles in a
typical branch $w_k$ of a Cayley tree is $\sim k/2$. Therefore it
takes $\sim k/2$ returns to the root for the domain of visited
sites to reach the $k^{\rm th}$ generation of the tree.

The median $m(n)$ of the duration of $n^{\rm th}$ return loop
satisfies
\[
\sum_{k>m(n)/2}^\infty\Pr[T_n-T_{n-1}=2k]\sim\sum_{k>m(n)/2}^\infty
\frac{1}{2^n\Gamma(1-2^{-n})p^{1-2^{-n}}k^{1+2^{-n}}}=\frac{1}{2}.
\]
Replacing the sum by an integral and solving the equation for
$m(n)$, we obtain
\[
m(n)\sim 2^{2^n}c,
\]
as $n\to\infty$. That is, with probability $\frac{1}{2}$, it takes
over $2^{2^n}c$ time units for the walker to complete the $n^{\rm
th}$ return path.

According to Corollary 3
\[
T_n-T_{n-1}\leq T_n\leq 2(T_n-T_{n-1}).
\]
Hence the median of $T_n$ is between $2^{2^n}c$ and $2^{2^n+1}c$
once $n$ is sufficiently large.

The $k^{\rm th}$ generation of the tree is reached with
probability $\frac{1}{2}$ at a time $t\sim c\, 2^{2^{k/2}}$.
Solving the equation $t=c\, 2^{2^{k/2}}$ for $k$, we obtain
\[
k\sim 2\log_2\log_2 t,
\]
as $t\to\infty.$

\hfill \rule{2mm}{2mm}

\section{Discussion and concluding remarks.} The analysis of the
previous sections can be generalised to the case of a Cayley tree
with the branching ratio $b>2$ at the expense of extra technical
efforts. Let the arrow directions at every site be numbered
counterclockwise $0,1,2,\ldots,b$ starting from the direction
towards the root. Let also $\Pr[X=k]=p_k$, $k=0,1,\ldots,b$ be the
initial distribution of arrow directions at every site of the
tree. Then we can associate a realization of a branching process
to every initial configuration of arrows as follows. From the
first generation on, a particle of the associated branching
process at a particular site of the tree has
$k\in\{0,1,\ldots,b\}$ descendants if the arrow at that site
points in the direction number $k$. The new particles are placed
immediately above the parent at the adjacent sites in the
directions $1,2,\ldots,k$.

The associated branching process is critical if
$\sum_{k=1}^{b}kp_k=1$. Already for $b=3$ the explicit formula for
the generating function of the total number of particles in the
branching process, $f(x)=\mbox{\boldmath $E$}x^Z$, becomes very
cumbersome. For $b>4$ we lose the luxury of explicit formulae
completely. Nevertheless, the results of Lemma 1 are not difficult
to derive for the case of general $b$. For the condensed phase the
formulae for differentiation of implicit functions yield
\begin{eqnarray}
&&m_1\equiv\mbox{\boldmath $E$}T_1=\frac{2}{1-\sum_{k=1}^bkp_k};\nonumber\\
&&\mbox{Var}\,T_1=\frac{4\sum_{k=0}^b(k-1)^2p_k}
{\left(1-\sum_{k=1}^bkp_k\right)^3}-\frac{4}{1-\sum_{k=1}^bkp_k}.\nonumber
\end{eqnarray}
The tail of the first-return probability, $P[T_1=2k]$, can be
described in terms of a positive solution, $x^*$, of the equation
\[
\sum_{k=2}^bp_k(k-1)x^k=p_0.
\]
Namely, see \cite{o49},
\[
\Pr[T_1=2k]\sim \sqrt{\frac{f(x^*)}{2\pi f''(x^*)}}\,
k^{-3/2}\left(\frac{f(x^*)}{x^*}\right)^k,
\]
as $k\to\infty$.

We see that the properties of the first-return time for $b>2$ are
qualitatively similar to the analogue results in the case of the
branching ratio 2. In a similar way all the conclusions of the
previous sections can be generalized to the case $b>2$, and the
generalization does not produce a novel behaviour.

Of course the Euler walk on a Cayley tree is only a toy version of
Euler walks on 2D or 3D lattices. Nevertheless we believe/hope
that some of the main features of the Euler walk described in this
paper are also present in finite-dimensional cases. In particular,
we believe that finite-dimensional walks also have the condensed
and the low-density phases, and a transition between them.

Martingales might prove to be also useful for investigation of the
finite-dimensional walks, but in what way and to what extent is
yet to be discovered. Some general properties of the growth of the
domain of visited sites on 2D lattices might be similar to those
found in the present paper. In particular, the drop of density
from 1 to 0 in 2D case might still be described by the error
function, cf.\ Eq.\ (\ref{dvs}). The relationship between the size
of domain and fluctuations of its boundary might still be the same
square-root law as in Hypothesis 1. It is possible to state a few
more similar hypothesis, however, the last one already sounds very
bold, and it might be dangerous to continue any further. In any
case, analytical investigation of the growth of domain of visited
sites for finite-dimensional lattices looks like a very tough
problem indeed.

Monte Carlo simulations for square lattices with {\em equally
likely\/} initial directions of arrows at every site were
conducted in the papers \cite{p96,p98}. The simulations show that
for 2-D square lattice the radius of the domain of visited sites,
$R(t)$, grows with time as $R(t)\sim c\,t^{1/3}$. It was also
conjectured that on a 3-D square lattice and in higher dimensions
we have a diffusive behaviour, $R(t)\sim c\,t^{1/2}$, because the
walker does not return to the cluster of visited sites frequently
enough. The last conjecture seems to imply that as the lattice
dimension tends to infinity the behaviour of the walker does not
become more and more similar to that of a walker on a Cayley tree.
However, on the basis of results obtained in this paper one can
put forward the following alternative interpretation of the
``diffusive" behaiviour of the walker on a 3-D lattice. It might
be the case that the Monte Carlo simulations for 3-D lattices were
simply conducted in the low-density phase where a compact domain
of visited sites is not formed. Changing the initial distribution
of arrows one can get into the condensed phase, where the radius
of the domain of visited sites grows, presumably, as $R(t)\sim
c_d\,t^{1/(d+1)}$, converging to the (Cayley tree) logarithmic
behaviour as the lattice dimension $d\to\infty$.

Something similar actually happens on Cayley trees as well. If we
take a Cayley tree with the branching ratio $b=2$, then the
equally likely initial distribution of arrows
$p_0=p_1=p_2=\frac{1}{3}$ puts the walker at the critical point
$p_1+2p_2=1$. While if we increase the branching ratio to $3$,
then the equally likely distribution $p_0=p_1=p_2=p_3=\frac{1}{4}$
corresponds to the low-density phase $p_1+2p_2+3p_4>1$.

It was already known that branching processes are relevant to and,
in fact, provide a mean-field description for some model of
self-organized criticality, see, e.g.\ \cite{a88,z95}. Although
branching process are also relevant to Euler walks, the latter
apparently belong to a somewhat different class of models, since
instead of fixed values for the standard set of critical
exponents, we have a whole spectrum of those. Indeed, instead of
the mean-field exponent $\tau=3/2$, describing the distribution of
the size of avalanches, we have the sequence $\tau_n=1+2^{-n}$,
$n=1,2,\ldots$, beginning with $3/2$.

Due to the infinite memory of the Euler walk it is difficult to
calculate the moments of the walker's location, $\mbox{\boldmath
$E$}x^k(t)$. It is a pity, since the second moment of the walker's
location for the simple random walk on, say, 2-D lattices,
$\mbox{\boldmath $E$}x^2(t)=ct$, is one of the main
characteristics of that random process. To partially fill this gap
we will extract some information on the behaviour of the second
moment from the results obtained in the previous sections. This
information might provide clues for explanation of a bizarre
behaviour of $\mbox{\boldmath $E$}x^2(t)$ for certain versions of
Euler walk on 2-D lattices \cite{pp06}. It is instructive to
compare at the same time the behaviour of the Euler walk on a
Cayley tree and the simple random walk on a 2-D square lattice.

Both the Euler walk on a Cayley tree in the condensed phase and
the 2-D simple random walk are recurrent. Here, however,
similarities end. While the expected return-time (and even the
variance) for the Euler walk is finite, the expected return-time
for 2-D random walk is infinite. As a consequence we have
monotonically increasing variance of the walker's location for the
2-D random walk, $\mbox{\boldmath $E$}x^2(t)= ct$. On a Cayley
tree the walker returns to the root at time instants $T_n$ with
$\mbox{\boldmath $E$}T_n<\infty$, $n=1,2,\ldots$, see the explicit
formulae in Corollary 2. If $\mbox{Var\,}T_1\ll(\mbox{\boldmath
$E$}T_1)^2$, then the returns to the root in the logarithmic scale
take place almost periodically, $\ln T_n\sim
n\ln(1+\frac{1}{2}\mbox{\boldmath $E$}T_1)$, as $n\to\infty$. On
the other hand, if $\mbox{Var\,}T_1\gg(\mbox{\boldmath
$E$}T_1)^2$, then the periodicity in the logarithmic scale turns
into chaotic behaviour without any visible pattern.

While in the latter case one can not rule out the monotonic
increase of $\mbox{\boldmath $E$}x^2(t)$, in the former case one
certainly has a nearly periodic vanishing of $\mbox{\boldmath
$E$}x^2(e^t)$. If the magnitudes of $\mbox{Var\,}T_1$ and
$(\mbox{\boldmath $E$}T_1)^2$ are comparable one should have an
intermediate situation with visible deviations in the shape of
$\mbox{\boldmath $E$}x^2(t)$ from a classic $ct^\gamma$ behaviour.
As we approach the critical point $q+2p=1$, the variance
\[
\mbox{Var\,}T_1\sim\frac{4(1-q)}{[1-(q+2p)]^3},
\]
grows faster than
\[
(\mbox{\boldmath $E$}T_1)^2=\frac{4}{[1-(q+2p)]^2},
\]
and we lose completely traces of the log-periodic behaviour.

\vspace{\abovedisplayskip}

{\bf Acknowledgements.\/} The author is grateful to V.\ B.\
Priezzhev for introduction to the subjects of self-organized
criticality and Euler walks.

\end{document}